\documentclass[conf]{new-aiaa}
\usepackage[utf8]{inputenc}
\usepackage{textcomp}
\usepackage{graphicx}
\usepackage{amsmath}
\usepackage[version=4]{mhchem}
\usepackage{siunitx}
\usepackage{longtable,tabularx}
\usepackage{caption}
\usepackage{subcaption}
\usepackage{comment}
\usepackage{tikz}
\usepackage{gensymb}
\usepackage{xcolor}

\usepackage{comment}
\usepackage{float}
\usepackage{longtable,tabularx}
\newcommand\crule[3][teal]{\textcolor{#1}{\rule{#2}{#3}}}
\newcommand\crules[3][orange]{\textcolor{#1}{\rule{#2}{#3}}}
\newcommand{\redstar}{\tikz[baseline=-0.1ex]\draw[fill=red, red, line width=0.2ex] (0,0) -- (0.02,0.06) -- (0.1,0.06) -- (0.04,0.12) -- (0.06,0.2) -- (0,0.14) -- (-0.06,0.2) -- (-0.04,0.12) -- (-0.1,0.06) -- (-0.02,0.06) -- cycle;}
\title{Investigation of Flight Conditions where Box-Wing Outperforms Mono-Wing Configurations for Small UAVs}

\newcommand{\tikzcircle}[2][red,fill=red]{\tikz[baseline=-0.5ex]\draw[#1,radius=#2] (0,0) circle ;}%

\newcommand{\vo}[1]{\boldsymbol{#1}}

\newcommand{\fig}[1]{Fig. \ref{fig:#1}}
\newcommand{\flab}[1]{\label{fig:#1}}

\author{Jasmine Jerry Aloor  \footnote{{Graduate student, Department of Aeronautics and Astronautics, Massachusetts Institute of Technology, USA (Work largely done as an undergraduate student, while at Indian Institute of Technology Kharagpur).}} }\affil{Department of Aeronautics and Astronautics, Massachusetts Institute of Technology, USA}  
\author{Bikalpa Bomjan Gurung \footnote{Graduate student, School of Engineering Technology, University of New South Wales, Australia (Work done largely as a graduate student while at Indian Institute of Technology Kharagpur).}}\affil{School of Engineering Technology, University of New South Wales, Australia.}
\author{Gauri Wadhwa\footnote{Graduate student, Department of Mechanical and Aerospace Engineering, Princeton University, USA (Work largely done as an undergraduate student, Department of Aerospace Engineering, Indian Institute of Technology Kharagpur).}}\affil{Department of Mechanical and Aerospace Engineering, Princeton University, USA.}
\author{Mohit Singh\footnote{Graduate student, Department of Engineering Cybernetics, Norwegian University of Science and Technology, Norway (Work largely done as an undergraduate student, while at Indian Institute of Technology Kharagpur).}}\affil{Department of Engineering Cybernetics, Norwegian University of Science and Technology, Norway.}
\author{Raktim Bhattacharya\footnote{Associate Professor, Department of Aerospace Engineering, Texas A\&M University, USA. Associate Fellow AIAA}} \affil{Department of Aerospace Engineering, Texas A\&M University, USA.}
\author{Sandeep Saha\footnote{Associate Professor, Department of Aerospace Engineering, Indian Institute of Technology Kharagpur, India. \newline
Corresponding Author: Sandeep Saha \newline Email address: ssaha@aero.iitkgp.ac.in}}
\affil{Department of Aerospace Engineering, Indian Institute of Technology Kharagpur, India.}
\begin{document}

\maketitle

\begin{abstract}
We investigate the aerodynamic efficiency and flight dynamics of mono-wing and box-wing configurations across various parameters, including aspect ratio, velocity, and lift requirements. We find that although mono-wing configurations exhibit superior aerodynamic efficiency in certain regimes, box-wing designs perform better in circumstances like high velocities and increased lift demands. Box-wing configurations also prove advantageous when induced drag is higher than friction drag due to their ability to suppress the tip vortices. Furthermore, while analyzing the flight dynamics, low aspect ratio box-wing configurations show improved gust tolerance and stability in longitudinal and lateral dynamics. However, no substantial difference in flight dynamics is observed between box-wing and mono-wing designs for high aspect ratio configurations. The findings underscore the importance of selecting the appropriate wing configuration based on specific performance requirements and operational conditions.
\end{abstract}

{\renewcommand\arraystretch{1.0}
\noindent\begin{longtable*}{@{}l @{\quad=\quad} l@{}}
$\alpha$ & angle of attack (deg, $\degree$)\\
$AR$ & aspect ratio\\
$b$ &span (m)\\
$c$ & root-chord length (m) \\
$C_D$ & drag coefficient of wing \\
$C_{D0}$ & zero-lift drag coefficient of wing \\
$C_{Df}$ & skin friction drag coefficient of wing \\
$C_{Dp}$ & pressure drag coefficient of wing \\
$C_d$ & drag coefficient of airfoil \\
$C_L$ & lift coefficient of wing \\
$C_l$ & lift coefficient of airfoil \\
$C_M$ & pitching moment coefficient of wing \\
$C_L/C_D$ & aerodynamic efficiency of a wing \\
$MAC$ & mean aerodynamic chord (m)\\
$Re$ & Reynolds Number (based on the root-chord and freestream velocity) \\
$S$ & wing area (m$^2$)
\end{longtable*}}

\section{Introduction}
\lettrine{O}{ver} the last decade, the study of Micro Aerial Vehicles (MAVs) and Unmanned Aerial Vehicles (UAVs) has experienced remarkable growth, primarily due to their extensive applications in both civil and military sectors \cite{dietrich2020,usenko2020tum,gago2020nano,iv2006assessment,vskrinjar2018application,luo2011multi,asadpour2016route,martinez2018state,dames2015active}. This can be attributed to their compact dimensions, maneuverability, and remote operation capabilities. Nevertheless, their small size and low operating altitude render them more susceptible to atmospheric disturbances or gusts \cite{Moschetta_MAV_aero}.

Several techniques have been employed to address this issue, including advanced actuation and control systems, although these come with added complexity and weight. Alternatively, designing wings with high stall angles, such as delta wing configurations, low aspect ratio wings, biplane wings, or box wing configurations, can help mitigate gust instability. Fixed-wing designs have better endurance and a low thrust-to-weight ratio, ideally suited for applications requiring relatively large payloads to be transported long distances or long-endurance missions \cite{boon2017comparison,keane2017small,cai2010brief}. Thus, a rising interest in the exploration of fixed-wing designs naturally prompts the comparison of the aerodynamic performance of non-conventional concepts \cite{letcher1972v,jansen2010effect, traub2009experimental,jansen2010aerostructural,hicken2010induced,yamazaki2014biplane} with their conventional counterparts. In this spirit, we compare the aerodynamic performance of box-wings to mono-wings for small UAVs to identify flight conditions where each exhibits superior performance compared to the other.
 
Box-wing configurations are fascinating since they offer several advantages over traditional wing designs, making them an attractive option for various aircraft applications. A `box-wing' is a joined-wing design (see Fig. \ref{fig:ModelImages}(c,d)) first proposed by Prandtl \cite{prandtl1924induced}, with equal lift distribution on the upper and lower wings and a symmetric lift distribution about the vertical joining surfaces (see Fig.2 of \citet{Ref17}). In particular, their structural efficiency is enhanced due to the closed-loop structure, which reduces weight and increases stiffness \cite{samuels1982structural,andrews2015wing}. This improvement in structural integrity allows for better overall performance and decreased material usage. Regarding aerodynamics, box-wing designs offer superior performance by reducing induced drag and minimizing wingtip vortices typically associated with conventional wings \cite{prandtl1924induced,russo2020box, Ref14, Ref13,demasi2016minimum}. This improves lift-to-drag ratios, increasing fuel efficiency and overall flight performance.  
Control authority is another area where box-wing designs excel. Their ability to facilitate direct lift and side force control  \cite{wolkovitch1986joined}  gives pilots greater authority over the aircraft's movement, enhancing responsiveness and stability, particularly in gusty or turbulent flight conditions. Box-wing configurations also exhibit superior stall characteristics compared to conventional wing designs \cite{boxwing_windTunnel, wolkovitch1986joined, bell2008design}. Their high stall angles and improved low-speed flight performance make them more resilient to atmospheric disturbances or gusts, enhancing aircraft safety and reliability. Lastly, the closed-loop structure of box-wing configurations can contribute to reduced noise emissions, resulting in a quieter flight experience for passengers and diminished noise pollution in the surrounding environment. These attributes make box-wing designs a promising and innovative alternative for future aircraft development.

One of the prominent works on box-wing configurations is the PrandtlPlane concept \cite{frediani2020conceptual,scardaoni2020prandtlplane}. This concept aims to achieve a more efficient and environmentally friendly aircraft design. The PPlane design has been studied for various applications, including passenger aircraft, cargo planes, and military aircraft. NASA's ERA (Environmentally Responsible Aviation) project explored box-wing designs to reduce fuel consumption, emissions, and noise pollution \cite{lockheed_NASA_ERA}. The Closed Wing Surface (C-WINGS) project explored the feasibility of implementing closed-wing configurations in future commercial aircraft. The project aimed to assess the aerodynamic, structural, and operational aspects of box-wing configurations. The findings from this project have contributed to a better understanding of the potential benefits and challenges associated with these unconventional aircraft designs \cite{verstraeten2009drag}.  Numerous academic studies have also been conducted to investigate the aerodynamic characteristics of box-wing configurations, focusing on areas such as lift-to-drag ratio, stall behavior, and control authority \cite{wolkovitch1986joined, Ref10, Ref16,demasi2014invariant, Ref18, Ref14, Ref15,garcia2016conceptual}. Computational Fluid Dynamics (CFD) simulations and wind tunnel tests have been widely used to analyze the performance of these designs under various flight conditions \cite{Ref17, cavallaro2016challenges}. 

While significant research has been conducted on box-wing aerodynamics for large-scale aircraft, studies focusing on MAVs remain limited. Nevertheless, the general advantages of box-wing configurations, such as improved stability, aerodynamic efficiency, and structural integrity, are still potentially applicable to MAVs. For instance, our previous experimental research \cite{boxwing_windTunnel} demonstrated enhanced stall characteristics in low-speed flight regimes for small UAVs and MAVs, which is beneficial for stall control under gusty and turbulent flight conditions. However, specific examples of box-wing designs tailored for Micro-Air Vehicles (MAVs) are scarce in the literature.

To address this gap, we investigate a particular box-wing configuration for MAV-class vehicles and compare the aerodynamics and flight mechanics performance to a conventional mono-wing configuration. Through computational simulations for aerodynamics and linear analyses for flight mechanics, this paper presents the potential benefits of box-wing designs in MAVs, focusing on aspects such as aerodynamic efficiency and gust tolerance. The comparison of box-wing to mono-wing configuration is not trivial because the outcome of the comparison depends on the parameters (for instance, Reynolds number, aspect ratio, lift generation) that have been held equal for the two wings being compared. Specifically, the main contributions of this paper are:
\begin{enumerate}
\item We comprehensively study the interdependencies between Reynolds number and aspect ratio on the aerodynamic properties of mono and box wing configurations for MAVs. Our study reveals that increasing the Reynolds number elicited a negligible effect on the lift coefficient while causing a marginal decrease in the drag coefficient, thus indicating a limited influence of the Reynolds number on parasitic drag. Furthermore, skin friction drag was observed to decrease in proportion to the fifth root of the Reynolds number. Upon maintaining the same Reynolds number, the box-wing configuration exhibited a higher skin friction drag than the mono-wing counterpart, which could be attributed to the additional friction drag introduced by the winglets. We also discovered that as the aspect ratio is decreased, a subsequent decline in the lift coefficient and an escalation in the drag coefficient occurs, with no discernible effect on the minimum drag coefficient. Notably, the induced drag coefficient increased with the diminishing aspect ratio, with a more substantial alteration observed in the case of the mono-wing model.
\item The aerodynamic characteristics of box-wing and mono-wing configurations for MAVs under diverse flight conditions are thoroughly assessed. Our findings demonstrate that, for equivalent aspect ratios ($AR$) and Reynolds numbers ($Re$), the box-wing configuration exhibits superior aerodynamic efficiency compared to its mono-wing counterpart due to reduced induced drag. Conversely, the mono-wing configuration presents enhanced aerodynamic efficiency when evaluated in comparable lifting areas, operational speeds, and aspect ratios under identical mission profiles. Moreover, for a given total lift condition, the maximum lift-to-drag ratio ({(L/D)}$_\mathrm{max}$) of the mono-wing configuration surpasses that of the box-wing at lower lift requirements. In comparison, the box-wing may exhibit greater aerodynamic efficiency at higher lift requirements. These critical insights underscore the necessity of meticulously considering many factors when selecting an optimal wing configuration to achieve peak aerodynamic performance.

\item A comprehensive nonlinear flight mechanics model for MAVs is developed, incorporating aerodynamics data from OpenVSP for various mono-wing and box-wing configurations. This model is then linearized around steady-level flight to facilitate further analysis. Eigenanalysis is conducted for both longitudinal and lateral modes to compare the performance of the two configurations. The most noticeable difference is observed in the phugoid modes, where box-wing configurations generally exhibit better damping but lower natural frequencies compared to their mono-wing counterparts. To assess the response of each configuration to gusts, time-domain simulations are performed, modeling gusts as step changes in velocity along the aircraft's $y$ and $z$ axes to simulate lateral and longitudinal responses, respectively. The results reveal that the box-wing configuration demonstrates superior convergence and damping in response to gusts, indicating its potential for enhanced stability and performance in gusty and turbulent flight conditions.
\end{enumerate}

By thoroughly examining the applicability of box-wing configurations for MAVs and comparing their performance with traditional mono-wing designs, this paper hopes to provide valuable insights into the ongoing development of innovative, efficient, and high-performance MAVs. Moreover, the results of this study have the potential to establish a foundation for subsequent research endeavors, optimization strategies, and practical implementation of box-wing configurations in the growing field of micro air vehicles.

The paper is arranged as follows. Section \ref{sec:companalysis} provides the details of the wing designs, meshing, boundary conditions, CFD analysis settings, and grid convergence tests. We present the validation studies of our computational analysis in Section \ref{sec:validation} against previous experimental and numerical simulation data \cite{cosyn, torres}. The results from the aerodynamics and stability analysis are presented in Section \ref{sec:results_ad} and \ref{sec:results_fm}, respectively. We conclude and outline future works and applications in Section \ref{sec:conclusion}.

\section{Computational Method}
\label{sec:companalysis}

\subsection{Geometrical and aerodynamic parameter selection of models}
\begin{figure}[thpb]
\includegraphics[width=\linewidth]{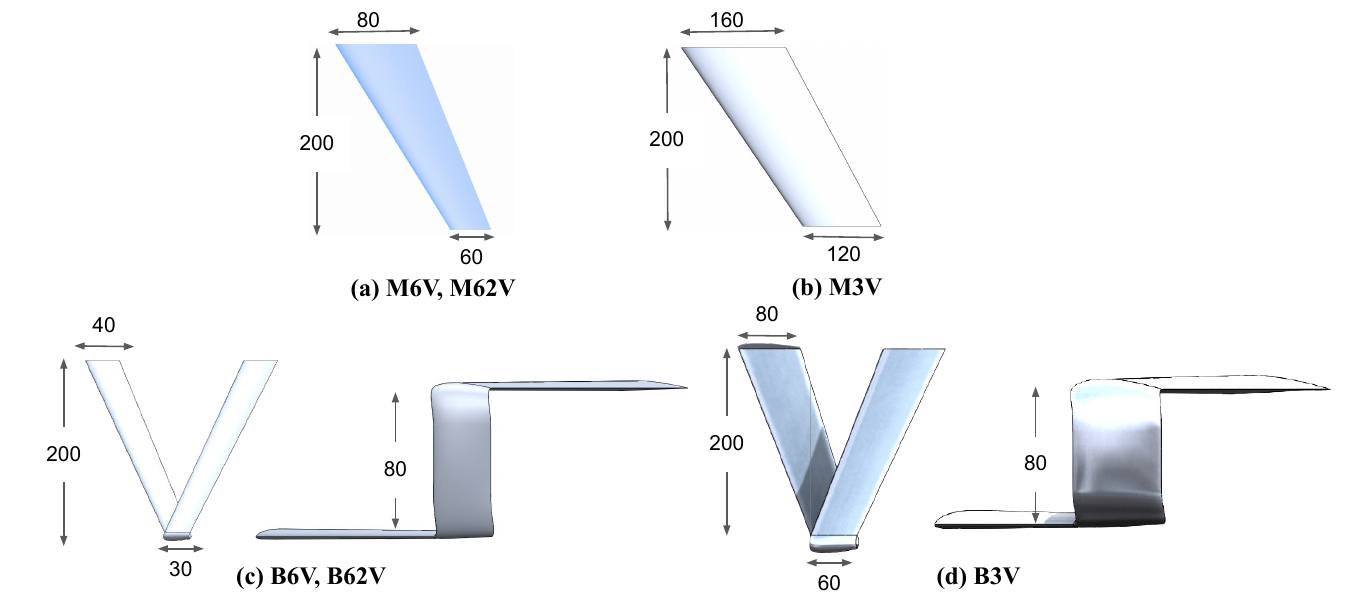}
\caption{Wing models for different Cases M6V-B3V (all dimensions in mm)}
\label{fig:ModelImages}
\end{figure}

A conventional swept-back mono-wing model was selected as a reference for this comparative study from our previous study \citep{boxwing_windTunnel}. The reference wing is designed using the Clark Y airfoil that has a root chord length ($c_r$) of 80 mm, semi-span ($b/2$) of 200 mm, a taper ratio of $0.75$ and sweep angle of $25\degree$. This reference mono-wing (hereafter referred to as M6V) is shown in Fig. \ref{fig:ModelImages}(a), and the associated parameters are summarized in Table \ref{tab:modeldetails}. Furthermore, the aerodynamic parameters for all the wings (M6V, M62V, M3V, B6V, B62V, and B3V) selected for our study are reported in Table \ref{tab:modeldetails} using a specific wing nomenclature. The wing nomenclature  M6V represents a mono-wing (M) with an aspect ratio $AR (b^2/S)$ of 6 cruising at V, \textit{i.e.} the design cruise speed of 22.14 m/s. A box-wing, B6V, corresponding to the same $AR$ and wingspan, was created for comparison against the mono-wing. Both the forward and aft wings of the box wing have half the chord length compared to the mono-wing so that the lifting surface area remains unchanged. The B6V  has area ratio (s$_{\mathrm{fwd}}$/S)= 0.5, stagger ${X_{1\xrightarrow{}2}}/b$ = 0.46 and gap h/b $=$ 0.2 \cite{Ref19}. M62V and B62V are wings with twice the cruise speed, V, chosen to study the effect of the Reynolds number variation. Additionally, two models, M3V and B3V, were set to have  $AR = 3$ while keeping their total surface area the same as that of the M6V and B6V models. This was done to investigate how changes in aspect ratio can affect aerodynamics and stability.  We summarize the dimensions of all models selected for our study in Fig. \ref{fig:ModelImages}

\begin{table*}[hbt!]
\centering
\caption{Model Geometry and Design Variables}
\label{tab:modeldetails}
\begin{tabular}{lllllllr}
\hline
{Model- Case }& Type & MAC  & Span     & \multicolumn{1}{l}{Total Area} & \multicolumn{1}{l}{$AR$} & \multicolumn{1}{l}{$Re$} & \multicolumn{1}{l}{Velocity} \\ 
& & \multicolumn{1}{c}{{(c = 70 mm) }} & \multicolumn{1}{c}{{(b = 400 mm)}} & \multicolumn{1}{c}{{(S = 0.028 $\mathrm{m}^2$)}} & \multicolumn{1}{c}{{($b^2/S$)}} & \multicolumn{1}{c}{{($ \times 10^5$)}}& \multicolumn{1}{c}{{(m/s)}}\\
\hline
M6V& Mono-wing           & c  & b   & S     & 6                        & 1.50  & 22.14    \\
M62V    & Mono-wing           & c  & b   & S      & 6                        & 3.00  & 44.28    \\
M3V & Mono-wing         & 2c & b   & 2S       & 3 & 3.00                 & 22.14     \\
\hline
B6V& Box-wing & 0.5c  & b   & S         & 6     & 0.75       & 22.14         \\
B62V & Box-wing & 0.5c  & b   & S             & 6   & 1.50    & 44.28 \\
B3V   & Box-wing            & c  & b   & 2S             & 3    & 1.50                 & 22.14 \\

\hline
\end{tabular}
\end{table*}

\subsection{Domain and boundary conditions}

The simulations were performed on the half-wing, taking into account the symmetry of the flow. A semi-spherical flow domain of radius ten times the root chord length ($c_r$) was used for the simulation. The spherical surface was set as far-field, the plane surface of the hemisphere as symmetry, and the wing surface as no-slip wall boundary conditions. 

\subsection{Flow Solver}
The finite volume-based simulations were performed using SU2 v7.0.6 \cite{palacios2013stanford} to solve the incompressible Reynolds-averaged Navier–Stokes (RANS) equation. The Menter's Shear Stress Transport, a two-equation turbulence model, was used with the RANS equation to analyze flows near walls and freestream accurately  \cite{MenterSST}. Upwind Flux Difference Splitting (FDS) scheme was selected for the convective numerical method, and the implicit Euler method was used for time discretization. The convergence criterion was set using the drag coefficient, achieving an asymptotic state with the absolute tolerance of $10^{-5}$.

\subsection{Mesh convergence study}
\begin{figure}[htpb]
\centering
\includegraphics[width=0.7\linewidth]{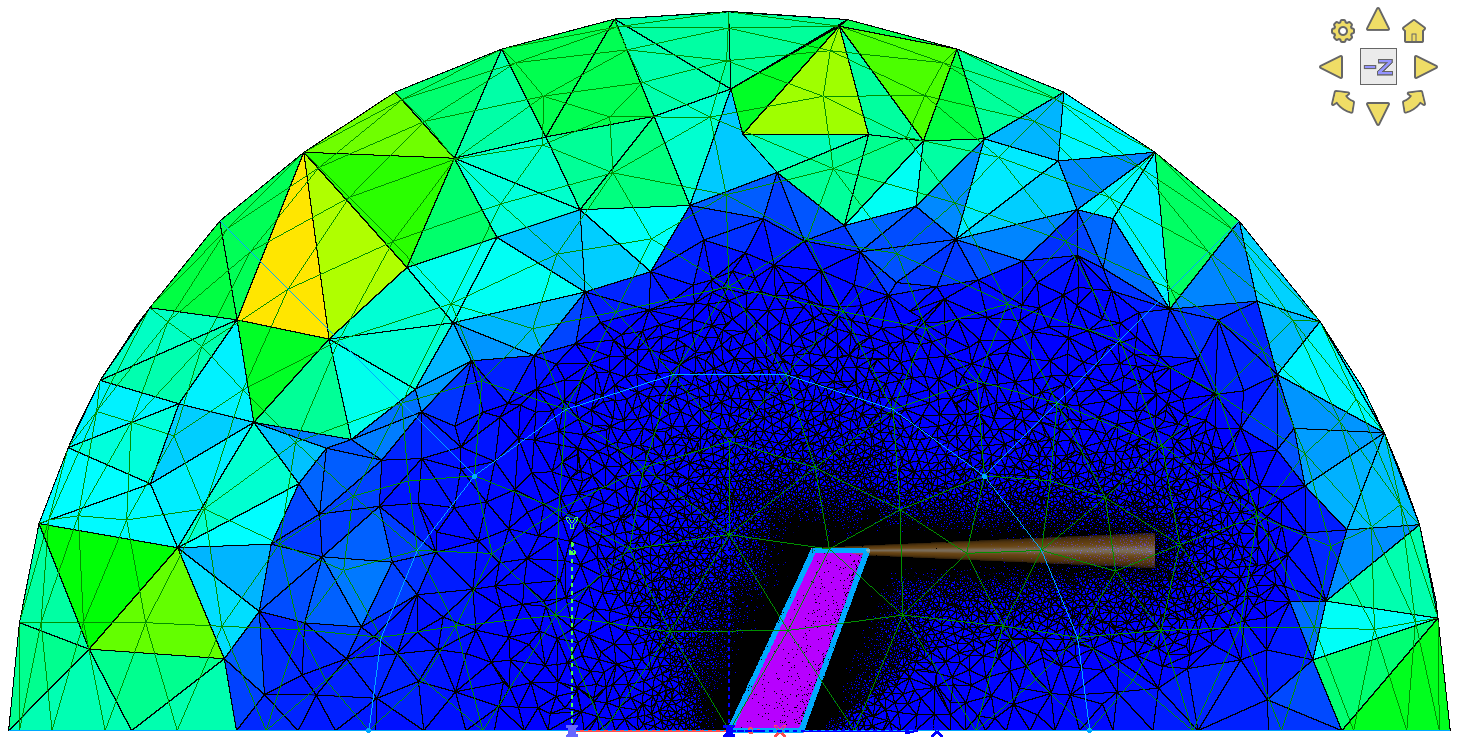}
\caption{Mesh with a source for the reference model M6V. A conical source is placed at the wingtip for local mesh refinement.}
\label{fig:source}
\end{figure}

A mesh convergence study was conducted on the reference model M6V at $\alpha = 5\degree$ to determine the appropriate mesh resolution. An unstructured surface mesh was generated for the wing surface, with a minimum edge length of 0.035 mm (5e-4$\times$c) on the wing's leading and trailing edges and an average edge length of 2 mm (0.028$\times$c) on the remaining surface. The surface mesh on the domain boundary had an average edge length of 40 mm (0.57$\times$c). Finally, an unstructured volume mesh was created from the surface mesh with a prism layer near the surface and tetrahedral on the remaining domain. The mesh had 30 inflation layers with a first layer height of 0.015 mm (2e-4$\times$c)  to get  $y^+ \sim 1$ at the wing surface and a growth rate of 1.1. The mesh was refined locally near the wing tip to resolve the wingtip vortices in the near-field region. To achieve this, a conical source was placed at the wingtip, with its axis extended up to 6 times the root chord $c_r$ and a grid spacing that varied from 0.035 mm (5e-4$\times$c) at the beginning to 3.5 mm (5e-2$\times$c) at the end. The effect of the source can be seen in Fig. \ref{fig:source}. This mesh is referred to as Type 2, and it serves as the baseline for further comparison.

Two additional meshes, Type 1 and Type 3, were generated to evaluate mesh convergence further. Type 1 mesh is a finer mesh with half the minimum and maximum edge length on the wing surface compared to the baseline mesh Type 2. On the other hand, Type 3 mesh is a coarser mesh with twice the minimum and maximum edge length on the wing surface compared to Type 2 mesh.   The total number of cells and grid points for all mesh types are shown in Table \ref{tab:gridConv}. Comparing coarse grid Type 1 to medium grid Type 2, we observe a lower value of $C_L(2.13\%)$ and $C_D(6.68\%)$ for medium mesh. Type 2 grid, when compared to fine grid Type 3, we note the further reduction in the $C_L$ (0.8\%) and $C_D$ (0.94\%). We select Type 2 settings for further analysis, considering the computation time and accuracy. The mesh is then created for all the models using Type 2 settings, and the total number of cells and grid points are summarized in Table \ref{tab:mesh}.

\begin{table}[hbt!]
    \centering
    \caption{Mesh convergence test for Case M6V- Mono-wing Design (Baseline) at $\alpha = 5\degree$}    \label{tab:gridConv}

    \begin{tabular}{lllllll}
    \hline\noalign{\smallskip}
         Grid Type & No. of Elements & $C_L$ & $C_D$ & $C_{Dp}$ & $C_{Df}$ & L/D \\ \hline\noalign{\smallskip}
        Type 1 & {4.29 million} & 0.5564 & 0.0389 & 0.0267 &  0.0123 & 14.27\\ 
        Type 2 & 9.7 million & 0.5445 & 0.0363 & 0.0239 & 0.0125 & 14.98 \\ 
        Type 3 & 19.84 million & 0.5402 & 0.0360 & 0.0235 & 0.0125 & 15.01 \\\noalign{\smallskip}\hline
    \end{tabular}
\end{table}

\begin{table*}[hbt!]
    \centering
    \caption{Mesh details for all cases}
    \label{tab:mesh}  
    \begin{tabular}{lllllll}
    \hline\noalign{\smallskip}
         Detail & M6V & M62V & M3V  & B6V & B62V & B3V \\ \hline\noalign{\smallskip}

        Total number of Cells (millions) & 9.7 & 9.7 & 14.7  & 17.6 & 17.6 & 26.9\\ 
        Total number of Points (millions) & 2.6 & 2.6 & 3.9  & 3.9 & 3.9 & 6.5 \\\noalign{\smallskip}\hline
    \end{tabular}
\end{table*}

\section{Validation Study}
\label{sec:validation}
We conduct a validation study for the comprehensiveness of our computational analysis using a three-dimensional simulation of a rectangular wing \cite{cosyn, torres}. A numerical study of LAR wings at low Reynolds number by \citet{cosyn} and the experimental on the same wing has been conducted by \citet{torres}. Simulation for the wing of $AR$ = 1 at $Re = 10^5$ was conducted to validate the numerical model used in our simulations.
The lift coefficient $C_L$ and drag coefficient $C_{D}$ obtained from our simulations are in quantitative agreement with the results of \citet{cosyn} and \citet{torres}. At higher angles of attack, there is a slight under-prediction of drag coefficient as compared to the experimental result \cite{torres}, but a slight overprediction of the numerical data of \cite{cosyn}. Similarly, the lift coefficient is slightly higher than that reported previously \cite{torres,cosyn} at high angles of attack. 
\begin{figure}[thpb]
\centering
\begin{subfigure}{0.285\linewidth}
\centering
\includegraphics[width=\textwidth]{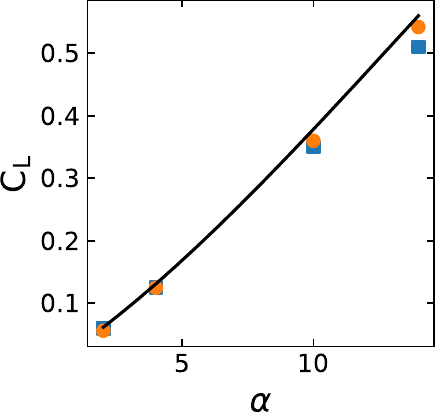}
\caption{Lift Coefficient}\label{fig:val2}
\end{subfigure}
\begin{subfigure}{0.3\linewidth}
\centering
\includegraphics[width=\textwidth]{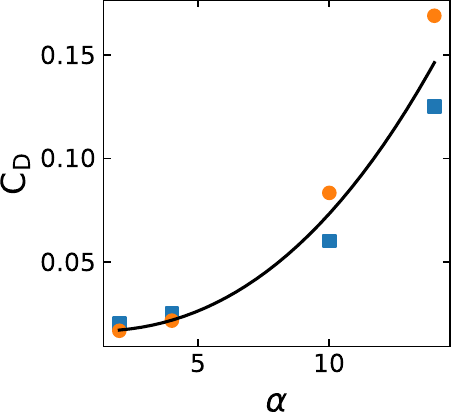}
\caption{Drag Coefficient}\label{fig:val1}
\end{subfigure}
\caption{Validation: Comparing aerodynamic data from literature (\tikzcircle[orange, fill=orange]{2pt}  Experiment (\citet{torres}), \crule{5pt}{5pt} numerical simulations (\citet{cosyn})) to present simulations }
\label{fig:val}
\end{figure}

\section{Aerodynamic Characteristics }
\label{sec:results_ad}

Understanding the relative performance of the mono- and box-wing models requires us to conduct a careful analysis wherein one parameter can be varied while the rest are held constant. However, interpreting such comparisons between mono and box wing models is far from trivial. 
For instance, a designer may find it practical to compare the M6V and B6V models with the same lifting surface area, aspect ratio, and operating speed despite the difference in Reynolds number (M6V at $1.5 \times 10^5$ and B6V at $0.75 \times 10^5$) due to the latter having half the chord length than the former. However, such a comparison provides little information about why one model performs better or worse than the other. It is unclear whether the difference is due to the models operating at different Reynolds numbers and, therefore, having different skin friction drag or due to differences in wing configuration, thus producing different induced drag. Therefore, we also need to compare the performance of wing designs from a pure aerodynamics perspective at the same aspect ratio and Reynolds number. This allows us to isolate the effects of wing configuration on aerodynamic performance. In this context, comparing the M6V and B62V models with the same aspect ratio ($AR$ = 6) and Reynolds number ($1.5\times 10^5$) would be an ideal choice. However, the cruise speed is not identical for the two wings, which makes deriving insight from the comparison hard for practical applications. The wing aspect ratio is another important parameter in determining aerodynamic performance, particularly in relation to induced drag. 
Thus, comparing wing configurations at different aspect ratios and cruise velocities is important for a more complete understanding.

Overall, the dependence of aerodynamic efficiency on aspect ratio and Reynolds number is complex when comparing box-wing and mono-wing designs. Thus, we first analyze the effect of Reynolds number and aspect ratio separately on the aerodynamic characteristics of the mono and box wing models. The comparison provides insight into the role of frictional and induced drag in determining the overall drag and, consequently, the aerodynamic efficiency of each wing model. The wings are then compared from various perspectives to gain a comprehensive understanding.

\subsection{Effect of Reynolds number}\label{sec:Re}

In this study, we investigated the effects of the Reynolds number on the aerodynamic force coefficients of two wing models, mono-wing and box-wing, at a constant aspect ratio  ($AR=6$). We analyzed two cases for the mono-wing, M6V, and M62V, which operated at $Re=1.5\times10^5$ and $Re=3.0\times10^5$, respectively, as well as two cases for the box-wing, B6V, and B62V, which corresponded to Reynolds numbers of $Re=0.75\times10^5$ and $Re=1.5\times10^5$, respectively see Table \ref{tab:modeldetails}.

\begin{figure}[h!]
\centering
\begin{subfigure}{0.32\linewidth}
\centering
\includegraphics[width=0.96\textwidth]{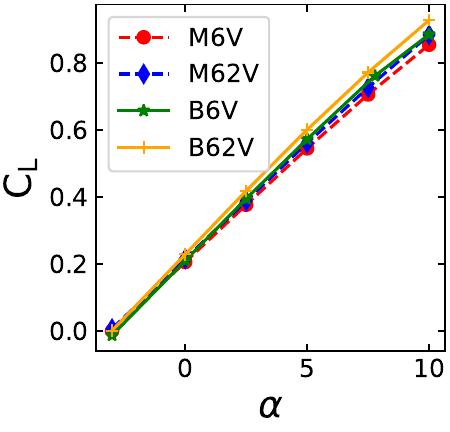}
\caption{$C_L$ vs. $\alpha$}\label{fig:recl}
\end{subfigure}
\begin{subfigure}{0.32\linewidth}
\centering
  \includegraphics[width=0.99\textwidth]{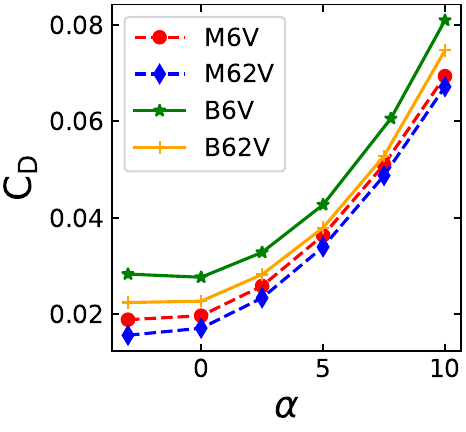}
\caption{$C_D$ vs $\alpha$}\label{fig:recd}
\end{subfigure}
\begin{subfigure}{0.34\linewidth}
\centering
  \includegraphics[width=0.99\textwidth]{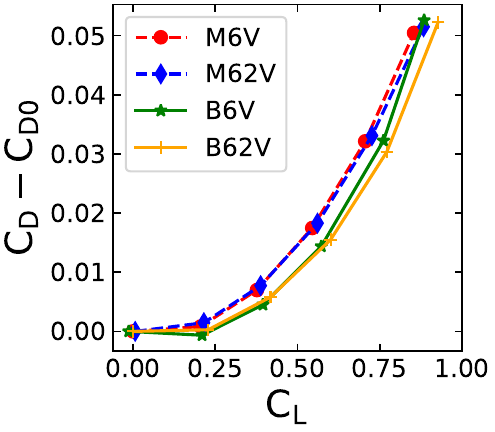}
\caption{($C_D$-$C_{D0}$)  vs $C_L$}\label{fig:cd_cdmincl}
\end{subfigure}
\caption{Dependence on Reynolds number: $Re$ for the different cases are M6V ($1.5 \times10^5$), M62V ($3\times 10^5$), B6V ($0.75\times 10^5$) and B62V ($1.5 \times10^5$). Each case has the same $AR$.}
\label{fig:re}
\end{figure}

Our results indicate that, for both types of wings, increasing the Reynolds number had an insignificant effect on the lift coefficient $C_L$. In contrast, the drag coefficient $C_D$ decreased slightly, as evidenced by Fig. \ref{fig:recl} and \ref{fig:recd}. Furthermore, we found no significant dependence of induced drag ($C_{D}-C_{D0}$) on the Reynolds number, as illustrated in Fig. \ref{fig:cd_cdmincl}, which confirmed that the effect of Reynolds number was limited to the parasite drag.

We compute the product of the fifth root of $Re$ and the value for the coefficient of skin friction drag ($C_{Df,0}$) based on $(1/7)^{th}$ power velocity profile law \cite{schlichting1979boundary}. We get $C_{Df,\mathrm{0}}\times Re^{1/5}\approx$ 0.1269 for the mono-wing models and $\mathrm{C_{Df,0}\times Re^{1/5}} \approx$ 0.1412 for the box-wing models. This suggests that $C_{Df,0}$ decreases as the fifth root of the Reynolds number. Furthermore, our observations indicate that at a given Reynolds number, the skin friction drag for the box-wing is higher than that of the mono-wing. Therefore, we speculate that the higher skin friction drag observed in the box-wing could be attributed to additional friction drag from the winglets.

\subsection{Effect of Aspect Ratio}\label{sec:AR}

\begin{figure}[thpb]
\centering
\begin{subfigure}{0.31\linewidth}
\centering
\includegraphics[width=0.99\textwidth]{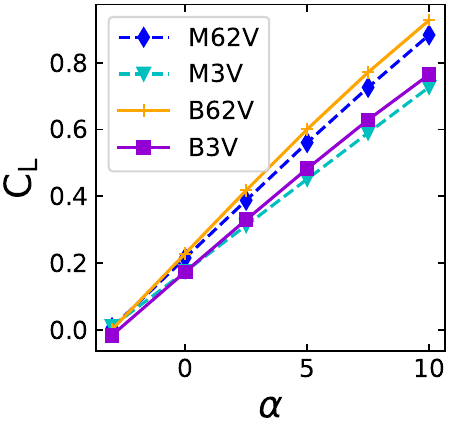}
\caption{$C_L$ vs. $\alpha$}\label{fig:arcl}
\end{subfigure}
\begin{subfigure}{0.32\linewidth}
\centering
  \includegraphics[width=0.99\textwidth]{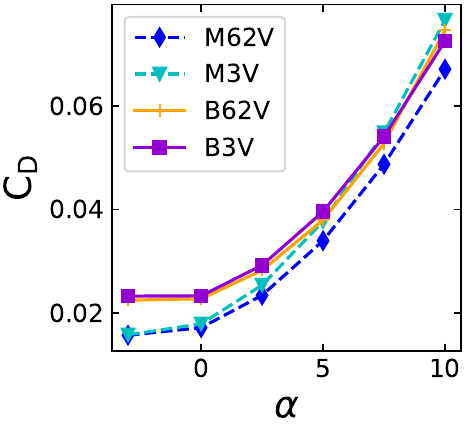}
\caption{$C_D$ vs $\alpha$}\label{fig:arcd}
\end{subfigure}
\begin{subfigure}{0.34\linewidth}
\centering
  \includegraphics[width=0.99\textwidth]{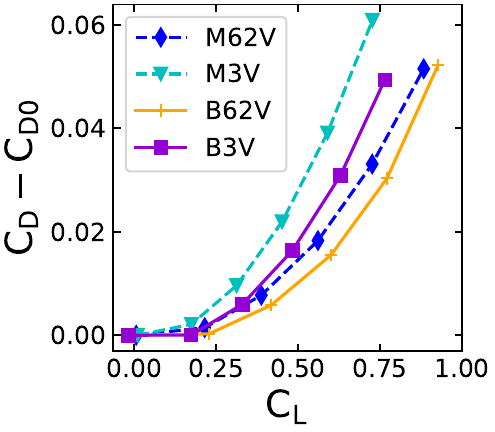}
\caption{($C_D$-$C_{D0}$)  vs $C_L$}\label{fig:cdcl}
\end{subfigure}

\caption{Dependence on Aspect Ratio: The $AR$ for the different cases are M62V (6), M3V (3), B62V (6), and B3V (3). M62V and M3V have $Re=3\times 10^5$. B62V and B3V have $Re = 1.5 \times 10^5$. }
\label{fig:ar}
\end{figure}

The effect of aspect ratio ($AR$) on the aerodynamic performance of mono-wing and box-wing models are studied with cases M62V($AR=6$), M3V($AR=3$) at the $Re=3.0\times10^5$, and B62V($AR=6$) and B3V($AR=3$) at the $Re=1.5\times10^5$.
Fig. \ref{fig:arcl}, shows that for both mono-wing and box-wing models, a decrease in $AR$ leads to a reduction in the lift coefficient slope ($C_{L\alpha}$) and subsequently, the lift coefficient ($C_L$). In Fig. \ref{fig:arcd} for the mono-wing and box-wing model, $C_D$ increases with the decrease in $AR$; however, the increase in $C_D$ is relatively small for the box-wing model. In Fig \ref{fig:cdcl}, it is observed that the difference between $C_D$ and $C_{D0}$, which represents the induced drag component dependent on $AR$ and increases as $AR$ decreases for both mono-wing and box-wing models. 
 
\begin{figure}[thpb]
\centering

\begin{subfigure}{0.4\linewidth}
\centering
  \includegraphics[width=0.99\textwidth]{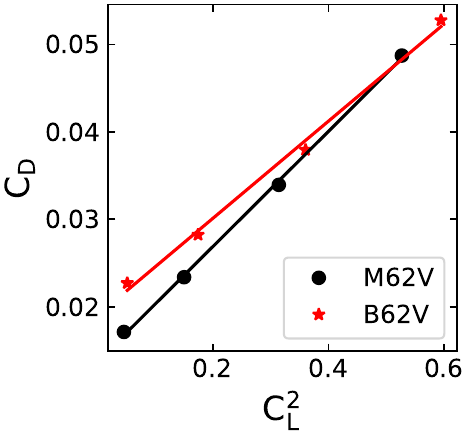}
\caption{M62V and B62V: (\redstar \hspace{1mm} B62V: $ C_D = 0.0556 \cdot {C_L}^2
+0.01898$), (\textbullet \hspace{1mm} M62V: $ C_D = 0.06593 \cdot {C_L}^2
+0.01368$) }\label{fig:cdvscl2_2bn6}
\end{subfigure}
 \hfill
\begin{subfigure}{0.41\linewidth}
\centering
  \includegraphics[width=0.99\textwidth]{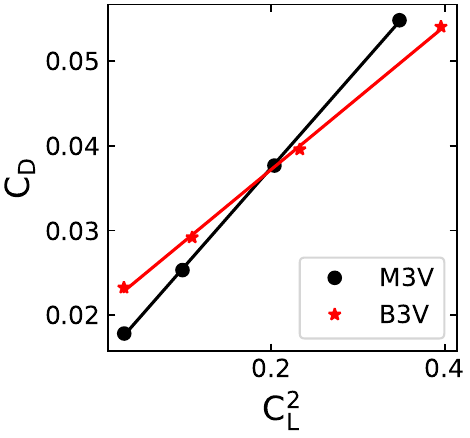}
\caption{M3V and B3V:  (\redstar \hspace{1mm} B3V: $ C_D = 0.08468 \cdot {C_L}^2
+0.02029$), (\textbullet \hspace{1mm} M3V: $ C_D = 0.11684 \cdot {C_L}^2
+0.01408$) }\label{fig:cdvscl2_3n4}
\end{subfigure}

\caption{Determination of Induced Drag Factor $k_{p+i}$}
\label{fig:cdvscl2}
\end{figure}

To understand the induced drag effect in detail, we compare the induced drag factor, $k_{p+i}$ for B62V, M62V, B3V, and M3V. We adopted the method used by \citet{traub2009analytic} to approximate the induced drag factor, where $C_D$ is plotted against $C_L^2$ and the slope of this polar gives $k_{p+i}$. The drag polar for the wing is then linearized as 
\begin{equation}
    C_D = C_{Do} + k_{p+i} {C_L}^2,
\end{equation}
where the first term corresponds to zero-lift drag, and the second term includes the contribution of induced drag and sectional-pressure drag. In Fig. \ref{fig:cdvscl2}, the slope of $C_D$ vs. $C_L^2$ for mono-wing models and box-wing models is calculated from the line equations derived using a line fitting method, and corresponding $k_{p+i}$ is shown in Table \ref{table:cfd_vsp2}

\begin{table}[hbt!]
    \centering
    \caption{induced drag coefficient and factor at $\alpha = 5\degree$}   \label{tab:induceddrag}

    \begin{tabular}{cllllll}
    
    \hline
        Model- Case & M62V & M3V & B62V & B3V \\ \hline
        $k_{p+i}$ & 0.0659 &  0.1168 & 0.0556 & 0.0847 \\ 
         \hline
            \end{tabular}
    \label{table:cfd_vsp2}
\end{table}

Table \ref{table:cfd_vsp2} reveals that a higher aspect ratio ($AR$) is associated with a lower induced drag coefficient $k_{p+i}$ for both the box-wing and mono-wing. Moreover, we observe that the induced drag coefficient for the mono-wing experiences a larger increment of 77$\%$ compared to 52$\%$ in the box-wing configuration, as the aspect ratio ($AR$) decreases from 6 to 3.

\subsection{Comparison of Box-wing and Mono-wing}
\label{subsec:bwing_mono_comp}

The comparison of box-wing and mono-wing models is not trivial and requires careful consideration of various factors. In this section, we place multiple constraints, for instance, constraining the aspect ratio and Reynolds number to be equal or cruise speed or lift to be equal, and we compare the performance of box-wing models to mono-wing models.

\subsubsection{Comparison under equal $AR$ and $Re$}

\begin{figure}[thpb]
\centering
\begin{subfigure}{0.44\linewidth}
\centering
\includegraphics[width=0.99\textwidth]{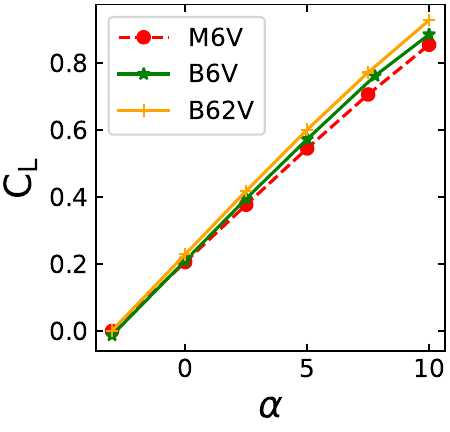}
\caption{$C_L$ vs. $\alpha$}\label{fig:comb_cl}
\end{subfigure}
\begin{subfigure}{0.46\linewidth}
\centering
  \includegraphics[width=0.99\textwidth]{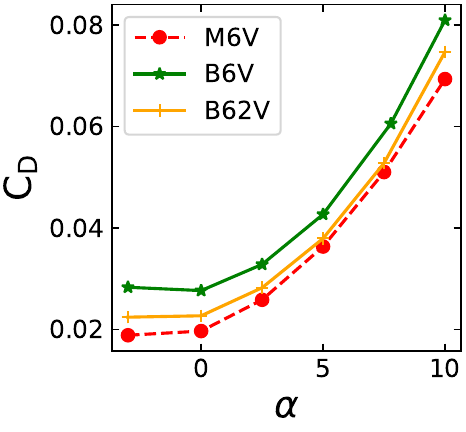}
\caption{$C_D$ vs $\alpha$}\label{fig:comb_cd}
\end{subfigure}
\begin{subfigure}{0.47\linewidth}
\centering
  \includegraphics[width=0.99\textwidth]{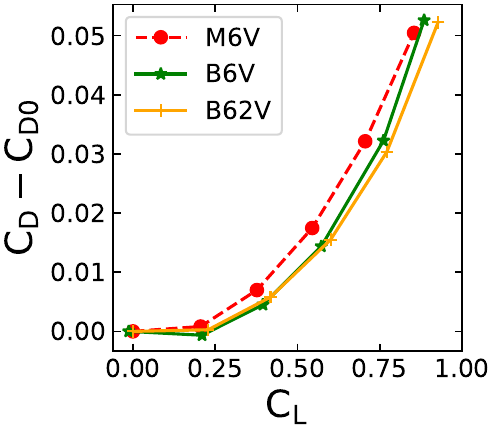}
\caption{($C_D$-$C_{D0}$)  vs $C_L$}\label{fig:cd_cdmin_cl}
\end{subfigure}
\begin{subfigure}{0.43\linewidth}
\centering
  \includegraphics[width=0.99\textwidth]{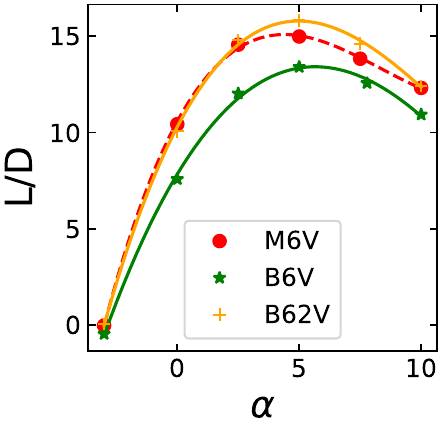}
\caption{L/D vs. $\alpha$}\label{fig:lbyd}
\end{subfigure}
\caption{Combined analysis of mono-wing and box-wing: Cases M6V and B62V have the same $Re = 1.5\times10^5$, Case B6V has $Re = 0.75\times10^5$ at the same $AR$ = 6}
\label{fig:comb}
\end{figure}

We compare the aerodynamic characteristics of the mono-wing M6V and box-wing B62V, which have equal $AR$ and $Re$. Our analysis reveals that the box-wing outperforms the mono-wing in terms of aerodynamic efficiency. Specifically, the data in Table \ref{tab:results} reveal that the B62V produces more lift (Fig. \ref{fig:comb_cl}) and an almost equal drag coefficient (Fig. \ref{fig:comb_cd}) to the M6V at $\alpha = 5\degree$. This difference can be attributed to the lower induced drag in the B62V, which results from the reduced downwash (Fig. \ref{fig:cd_cdmin_cl}), leading to a higher Lift to Drag ($L/D$) ratio for the B62V compared to the M6V (Fig. \ref{fig:lbyd}).

\begin{table}[hbtp]
    \centering
    \caption{Lift and Drag Coefficient of M6V, B6V, B62V at $\alpha = 5\degree$}
    \label{tab:results}  
    \begin{tabular}{lllllll}
    \hline\noalign{\smallskip}
        Case  & M6V & B6V & B62V \\ \hline\noalign{\smallskip}
        $C_L$ & 0.5445 & 0.5722 & 0.6005\\ 
        $C_D$ & 0.0363 &  0.0427 & 0.0379 \\
        $C_L/C_D$ & 14.9846 & 13.4033 & 16.5252        \\\noalign{\smallskip}\hline    \end{tabular}
\end{table}

\subsubsection{Comparison at same cruise condition}

In this study, we conducted a practical comparison between two wing models - the box-wing B6V and the mono-wing M6V - with similar lifting areas, operating speeds, and $AR$. Our goal was to observe and compare their aerodynamic characteristics while keeping the mission profile constant. The results showed that the B6V had lower aerodynamic efficiency compared to the M6V (Table \ref{tab:results}). Further analysis revealed that the B6V had a slightly higher lift coefficient slope, $C_{L\alpha}$, (Fig. \ref{fig:comb_cl}) but greater drag $C_{D0}$ (Fig. \ref{fig:comb_cd}), resulting in a lower $L/D$ compared to the M6V (Fig. \ref{fig:lbyd}).

We note that the B6V had a lower $Re$ than the M6V, which prompts the discussion of the contribution of individual drag components for each case (Section \ref{sec:Re}). At lower $Re$, the frictional drag becomes more dominant than the pressure drag, which contributes to the overall increase in drag. Additionally, the winglet surface present in the box-wing adds to the overall drag as a parasitic drag component. However, comparing the induced drag $C_D-C_{D0}$ at the same $C_L$ (Fig. \ref{fig:cd_cdmin_cl}) revealed that B6V had slightly lower induced drag compared to the M6V. This may be partially attributed to the reduced induced drag coefficient (Section \ref{sec:AR}) due to increased $b/c$ for each individual wing (using the conventional $AR$ definition).

Overall, the combined effect of the above components caused a significant increase in total drag with a small increase in lift, resulting in lower efficiency for the B6V (Fig. \ref{fig:lbyd}).

\subsubsection{Comparison under equal lift generation constraint}

\begin{figure}[thpb]
\centering
\begin{subfigure}{0.99\linewidth}
\centering
\includegraphics[width=0.99\textwidth]{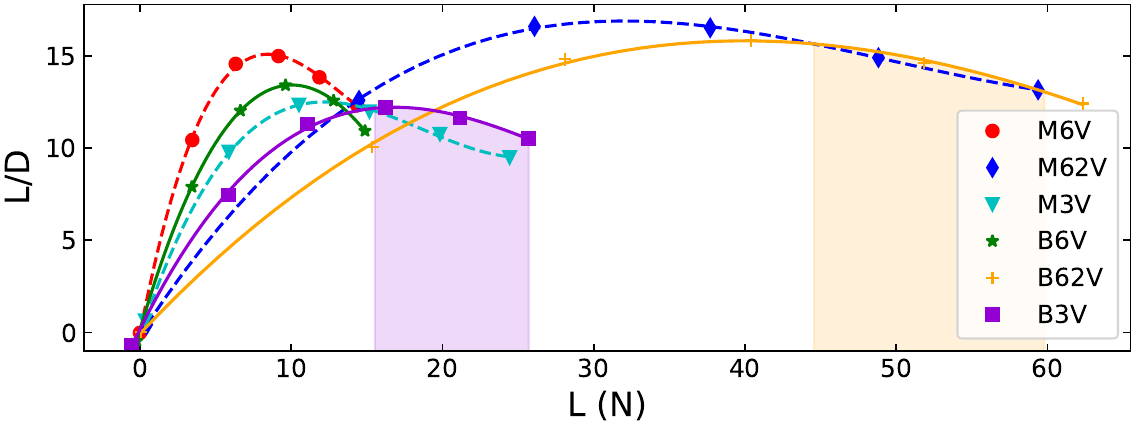}
\caption{L/D vs. Lift}\label{fig:ldvsl}
\end{subfigure}
\caption{Comparison under equal lift generation constraint (All models). Box-wing outperforms mono-wing in the shaded regions.}
\label{fig:liftindep}
\end{figure}

We finally compare the aerodynamic efficiency of all the wing models with respect to the lift generated in Fig. \ref{fig:liftindep}. The comparison reveals a few features of mono and box-wings that were not elucidated in the previous sections. If we compare the aerodynamic efficiency of the two wings with equal aspect ratio and equal cruise velocity, for instance, M6V and B6V, we observe that the mono wing M6V clearly outperforms the box wing B6V over the entire range of lift considered. However, the conclusion that monowing performs better than the box-wing cannot be generalized without adding a caveat. For instance, if the cruise velocity is doubled, keeping the aspect ratio fixed at 6, then we observe that B62V performs better than M62V for higher lifts and vice-versa for lower lifts. Hence, there exists a critical lift for which there is a switchover between the aerodynamic efficiency of the box-wing and the mono-wing. Alternatively, if we reduce the aspect ratio to 3, keeping cruise velocity fixed at V, the wing M3V performs better for low lifts, while B3V is more efficient at high lifts. The critical lift at which the switchover occurs is lowered as the aspect ratio is reduced. Hence, we conclude that the aerodynamic efficiency of the mono and box wings depends strongly on both aspect ratio and cruise velocity.

\subsection{Flight Condition where Box-Wing is Superior to Mono-Wing}

\begin{figure}[thpb]
\centering

\begin{subfigure}{0.45\linewidth}
\centering
\includegraphics[width=0.99\textwidth]{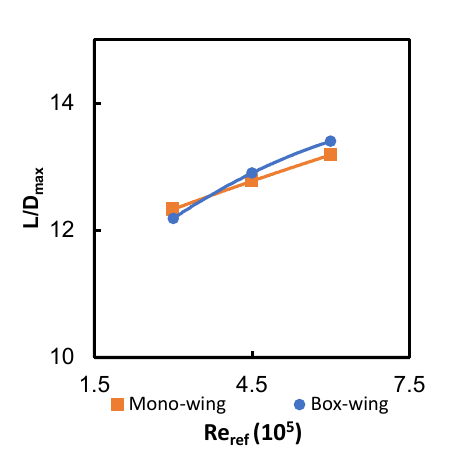}
\caption{$L/D_{\textrm{max}}$ vs $Re_{\textrm{ref}}$ at AR=3 \quad (\crules{5pt}{5pt} Mono $L/D_{\textrm{max}}$ = -0.007{$x^2$} + 0.3487$x$ + 11.35, \hspace{1mm} $R^2$ = 1), (\tikzcircle[blue, fill=blue]{2pt} Box $L/D_{\textrm{max}}$ = -0.0469{$x^2$} + 0.8276$x$ + 10.129, \hspace{1mm} $R^2$ = 1)}\label{fig:Crossover_AR3_r}
\end{subfigure}
\hfill
\begin{subfigure}{0.45\linewidth}
\centering
  \includegraphics[width=0.99\textwidth]{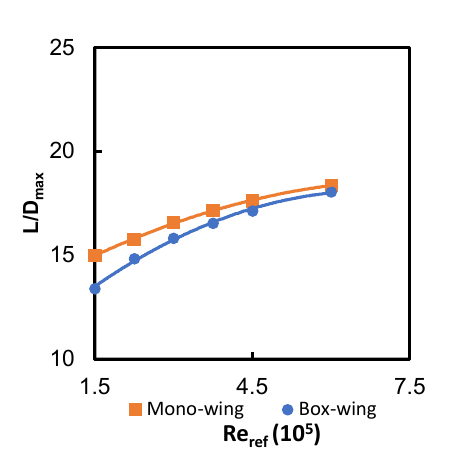}
\caption{$L/D_{\textrm{max}}$ vs $Re_{\textrm{ref}}$  at AR=6  \quad (\crules{5pt}{5pt} Mono $L/D_{\textrm{max}}$ = -0.094{$x^2$} + 1.4568$x$ + 13.012, \hspace{1mm} $R^2$ = 0.999), (\tikzcircle[blue, fill=blue]{2pt} Box $L/D_{\textrm{max}}$ = -0.1649{$x^2$} + 2.2386$x$ + 10.52, \hspace{1mm} $R^2$ = 0.9963)}\label{fig:Crossover_AR6_r}
\end{subfigure}

\caption{Comparing $L/D_{\textrm{max}}$ at different AR}
\label{fig:Crossover}
\end{figure}

With the aim of investigating whether there exist distinct conditions in which a wing design exhibits superior performance, we compare the maximum aerodynamic efficiency ($L/D_{\textrm{max}}$) of mono-wing and box-wing designs at various speeds and aspect ratios, while maintaining the same lifting area. The $L/D_{\textrm{max}}$ values are examined at several cruising velocities, specifically at velocities V, 1.5V, and 2V for aspect ratio 3 wings, and V, 1.5V, 2V, 2.5V, 3V, and 4V for aspect ratio 6 wings (see Fig. \ref{fig:Crossover_AR3_r} and \ref{fig:Crossover_AR6_r}). In the figure, $Re_{\textrm{ref}}$ is the Reynolds number based on the cruise speeds and the total mean aerodynamic chord length (sum of forward and rear wing MAC for box-wing) of the wing. Such a representation is important from a designer's perspective since it allows for the comparison of two equivalent wings with the same lifting surface area. A curve-fitting process has been applied to both figures to identify the trend line passing through all data points. The results indicate that as the Reynolds number ($Re_{\textrm{ref}}$) increases, the maximum aerodynamic efficiency ($L/D_{\textrm{max}}$) also increases for all the wing models. This trend can be explained by our previous observations in Section \ref{sec:Re}, where it was shown that the skin friction drag coefficient decreases as the cruise speed and Reynolds number increase, leading to a reduction in the overall drag coefficient while maintaining the same lift coefficient.

For the wings with an aspect ratio of 3, a point of crossover for the maximum $L/D_{\textrm{max}}$ in wing type occurs at a Reynolds number $Re_{\textrm{ref}}$ ~ $4\times10^5$, as depicted in Fig. \ref{fig:Crossover_AR3_r}. Beyond this Reynolds number, the box-wing design starts outperforming the mono-wing design. The specific Reynolds number at which this crossover occurs can be seen as the point where the box-wing design's advantage in reducing induced drag outweighs the skin-friction drag increment from the wing surface. A similar trend is observed for the wings with an aspect ratio of 6. However, the crossover is expected to occur at a higher cruise speed and Reynolds number compared to the wings with a lower aspect ratio of 3. This can be explained by the fact that at low aspect ratios, the box wings are more efficient in reducing the induced drag than at higher aspect ratios, as discussed in Section \ref{sec:AR}. As a result, the transition of better performance of the box-wing in comparison to the mono-wing happens at lower Reynolds numbers for smaller aspect ratios compared to higher aspect ratios.

\section{Flight Mechanics}
\label{sec:results_fm}

An aircraft's stability analysis provides information on how well it can fly and how smoothly it can be controlled. An aircraft in the air is subject to various disturbances, such as turbulence, wind gusts, or wind gradients. Good stability properties (handling qualities) enable the aircraft to successfully fly its mission and maneuver over a wide range of velocities. Knowing how an airplane responds to uncertain stimuli is essential in choosing a particular aircraft configuration.  In the following subsections, we present the flight mechanics and performance of the mono-wing and box-wing models and discuss the results.

\subsection{Full Aerodynamics Model}

\begin{figure}[htpb]
\centering
\begin{subfigure}{0.325\linewidth}
\centering
\includegraphics[width=0.99\textwidth]{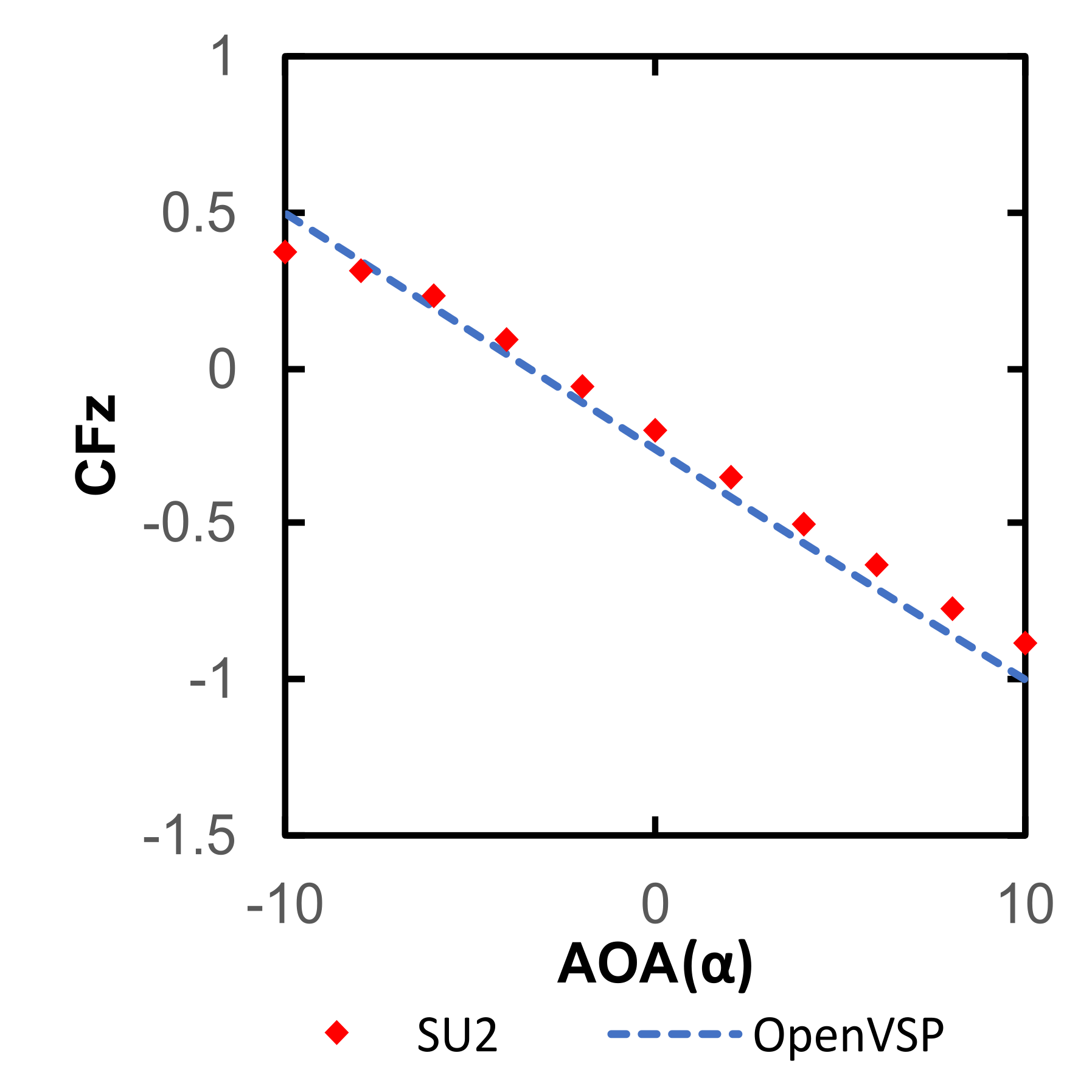}
\caption{}\label{fig:cfz_vs_a}
\end{subfigure}
\begin{subfigure}{0.325\linewidth}
\centering
\includegraphics[width=0.99\textwidth]{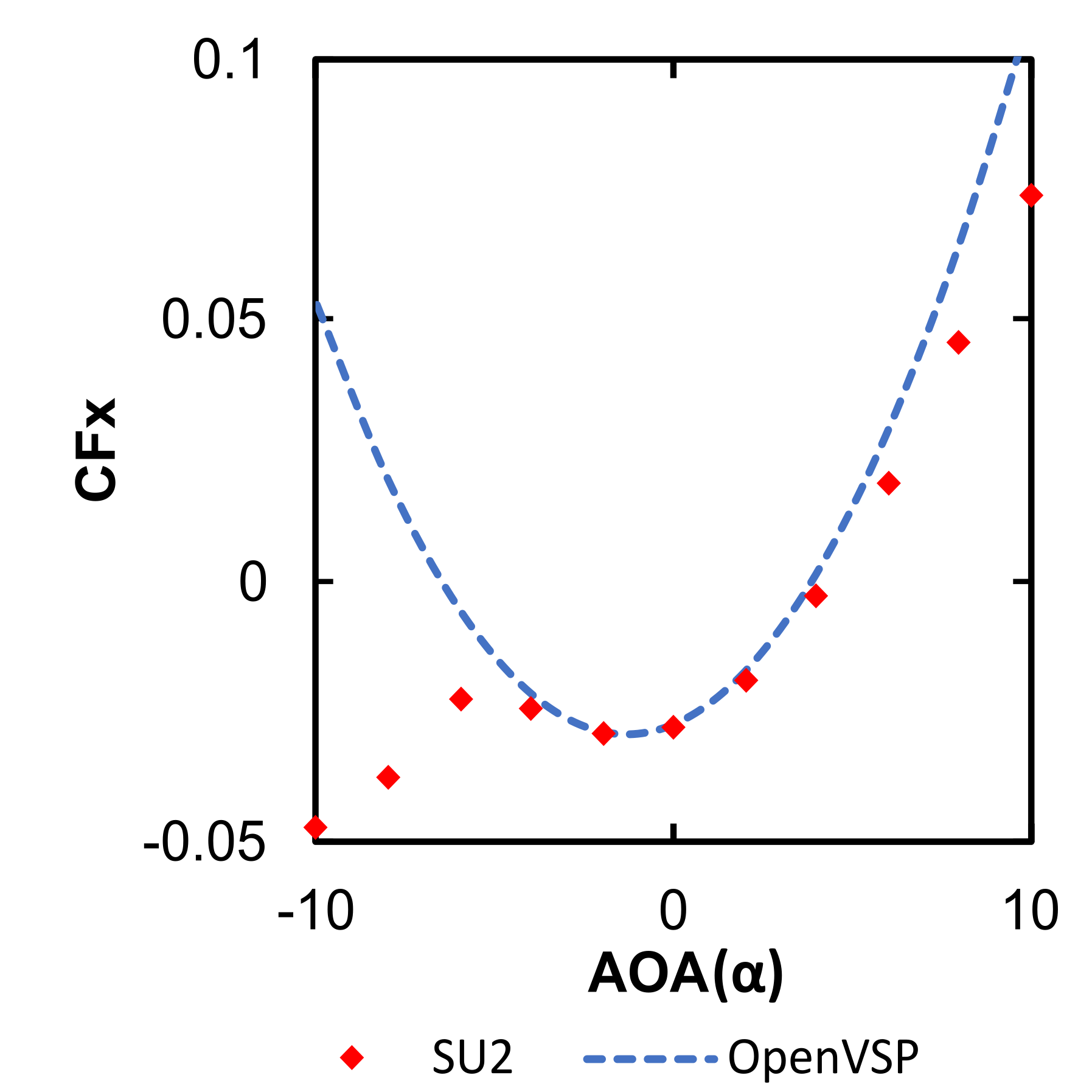}
\caption{}\label{fig:cfx_vs_a}
\end{subfigure}
\begin{subfigure}{0.325\linewidth}
\centering
\includegraphics[width=0.99\textwidth]{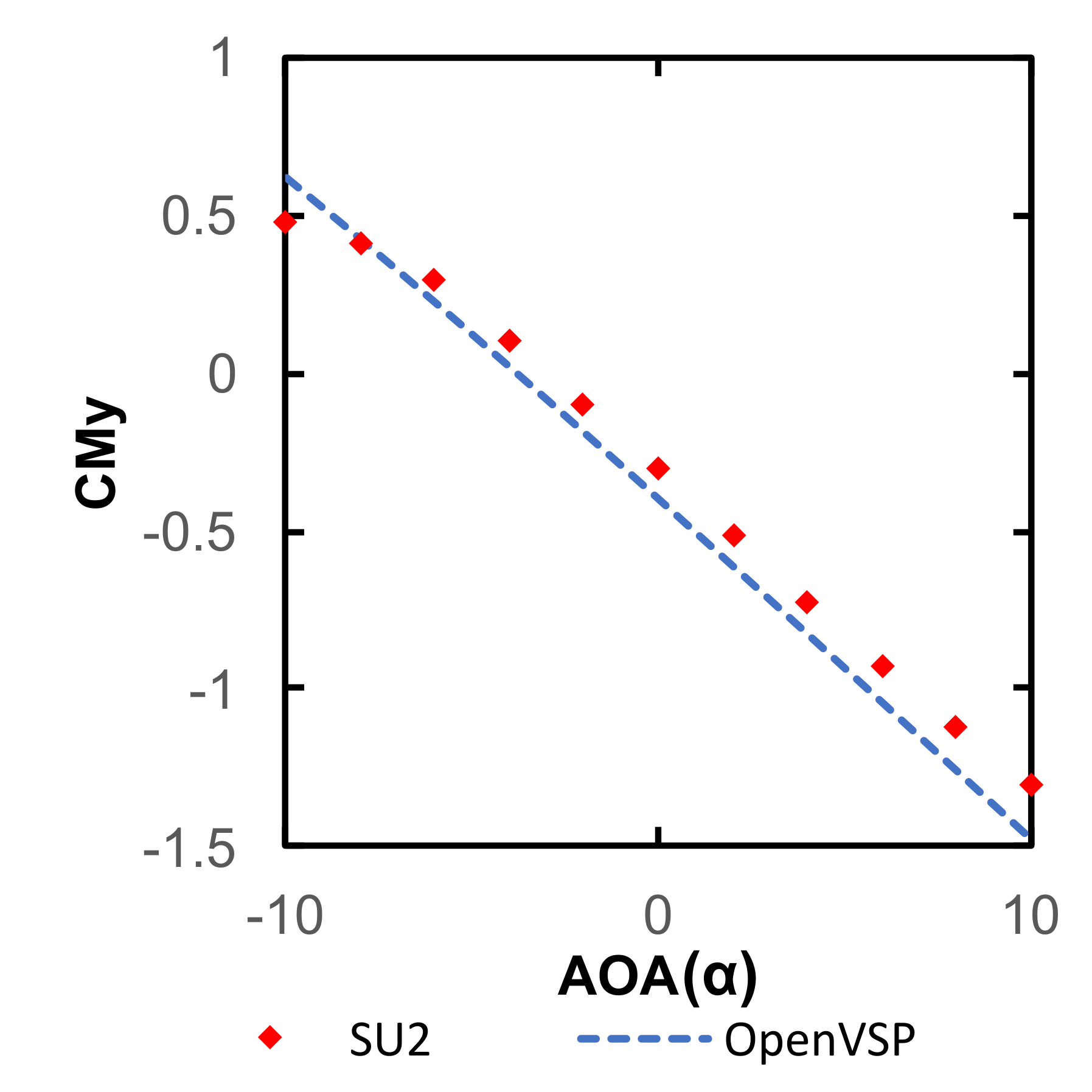}
\caption{}\label{fig:cmy_vs_a}
\end{subfigure}
\caption{$C_{Fz}$, $C_{Fx}$, $C_{My,LE}$. $\alpha$ for B6V model with SU2 and openVSP at $\beta = 0$}
\label{fig:calpha}       
\end{figure}

For the dynamic stability analysis, the low-fidelity Vortex Lattice Method (VLM) in OpenVSP was compared against the high-fidelity simulation in SU2. The aerodynamic coefficients for the B6V model computed with OpenVSP showed close agreement with SU2 at $\alpha >-4$ (Fig \ref{fig:calpha}). The deviation in $CF_x$ at $\alpha < -4 $ can be attributed to flow separation at these angles of attack, which VLM cannot simulate. Because $\alpha \approx  0$ is the region where dynamic analysis will be performed, OpenVSP was chosen to compute the aerodynamic coefficients and stability derivatives for the remaining models due to its low computational requirements. The force and moment coefficients for all models are shown in \ref{App1}.

\subsection{Rigidbody Dynamics}
The aerodynamics data from the above-described configurations were incorporated in a full nonlinear flight dynamics model in MATLAB/Simulink \cite{matlab}. The states of the models are the usual $$\vo{x} = \left[\begin{array}{cccccccccccc} x & y & z & u & v & w & \phi & \theta & \psi & p & q & r\end{array}\right]^T,$$
where $(x,y,z)$ are positions in the inertial frame, $(u,v,w)$ are velocities in the body-fixed frame, $(\phi,\theta,\psi)$ are the Euler angles and $(p,q,r)$ are the angular velocities in the body-fixed frame. We do not consider any control surfaces in the aircraft explicitly but introduce external forces and moments in the body-fixed frame as \textit{synthetic} control variables that can be realized in practice with suitable actuator architecture. This abstraction was necessary to trim the aircraft at a given flight condition. For our study, we trimmed the various aircraft configurations for steady-level flight at $22.14$ $m/s$. The trim states and controls were obtained using nonlinear least-squares optimization with constraints on state variables due to aerodynamics.

\subsection{Time Response}
We obtained linear models for the six vehicle configurations and investigated differences in natural frequencies and damping for longitudinal and lateral modes, which are summarized next.

\begin{figure}[h!]
\begin{subfigure}{0.5\textwidth}
\includegraphics[width=\textwidth]{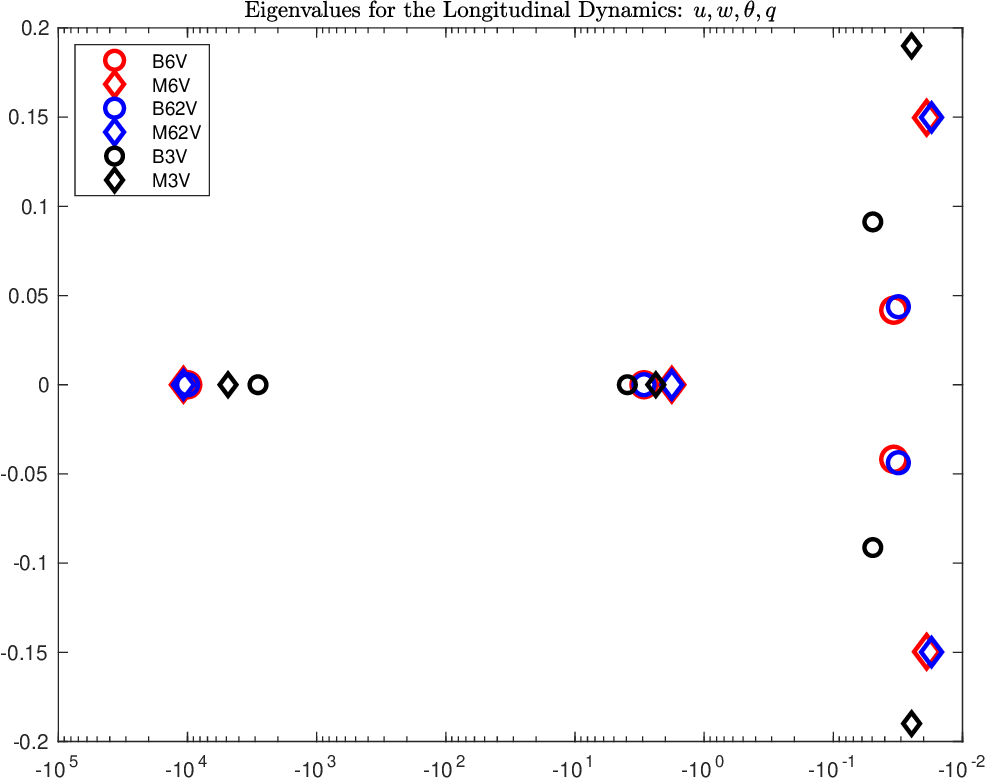}
\caption{Eigen values of the longitudinal dynamics.}
\flab{longi_evs}
\end{subfigure}
\begin{subfigure}{0.5\textwidth}
\includegraphics[width=\textwidth]{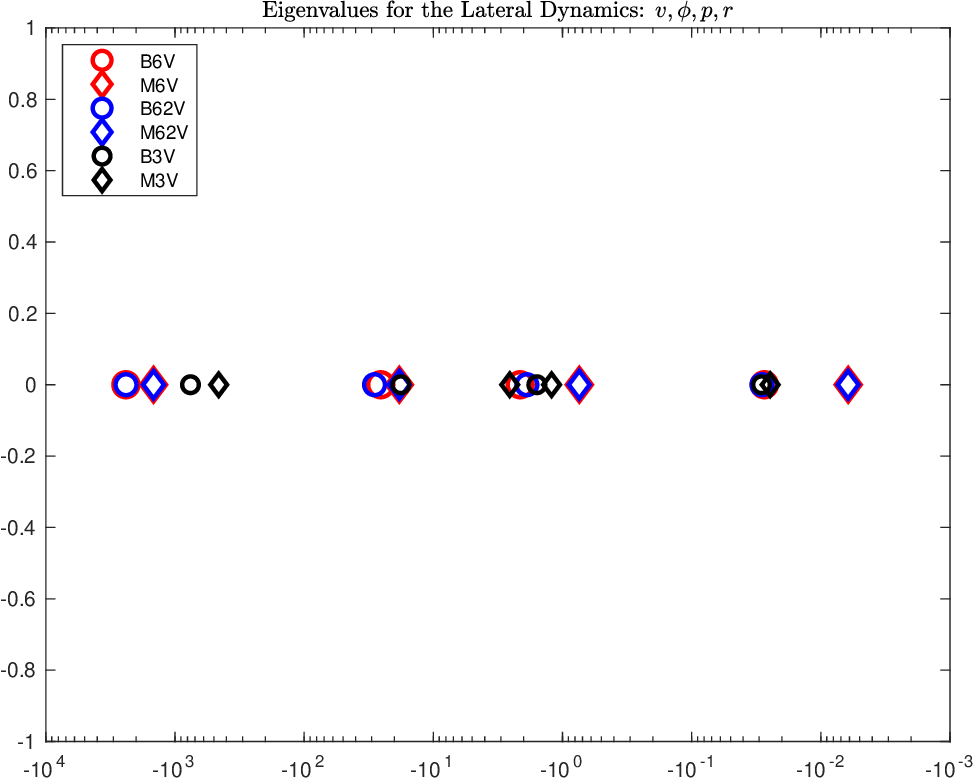}
\caption{Eigen values of the lateral dynamics.}
\flab{lat_evs}
\end{subfigure}
\caption{Comparison of the eigenvalues for various configurations.}
\end{figure}

\fig{longi_evs}, shows the eigenvalues for the longitudinal dynamics with states $u,w,\theta,q$. We observe that the eigenvalues of the short periods are all real, with a significant difference in the natural frequencies for all the configurations considered here. Also, there is not much difference between 6V and 62V configurations. The 3V configurations are marginally different, especially for the faster short-period mode. The difference between the configurations is pronounced in the phugoid modes. In general, the box-wing configurations have better damping but lower natural frequency than the corresponding mono-wing configurations. Therefore, we expect the mono-wings to be more responsive with higher oscillations than the box-wings. This behavior is observable in the step gust responses shown in \fig{longi_step}. We modeled the gust as a step perturbation in $w$ and simulated the time response shown in \fig{longi_step}. We observe faster decay with lower oscillations in the box-wing's response due to higher damping. We also observe a relatively shorter rise time in the mono-wing's response due to its higher natural frequency.

\begin{figure}[h!]
\begin{subfigure}{\textwidth}
\centering
\includegraphics[width=0.7\textwidth]{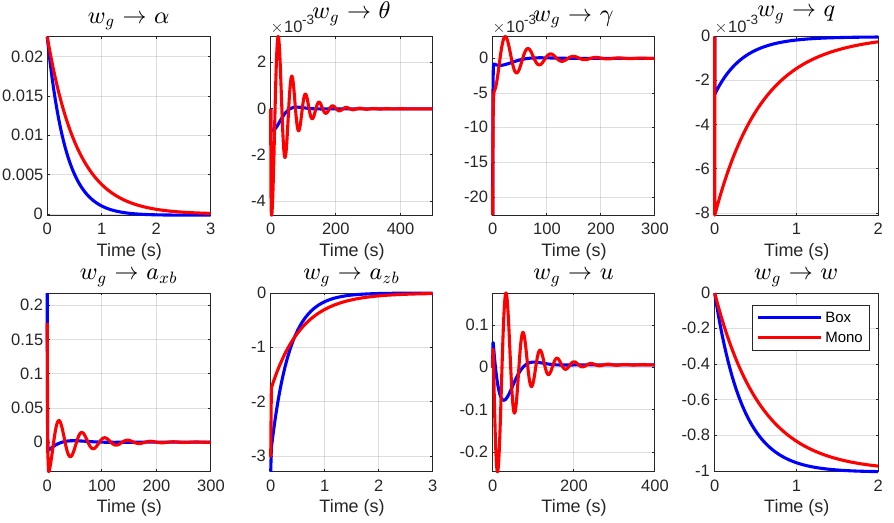}

\caption{B6V(blue), M6V(red).}
\flab{longi_step1}
\end{subfigure}\\
\begin{subfigure}{\textwidth}
\centering

\includegraphics[width=0.7\textwidth]{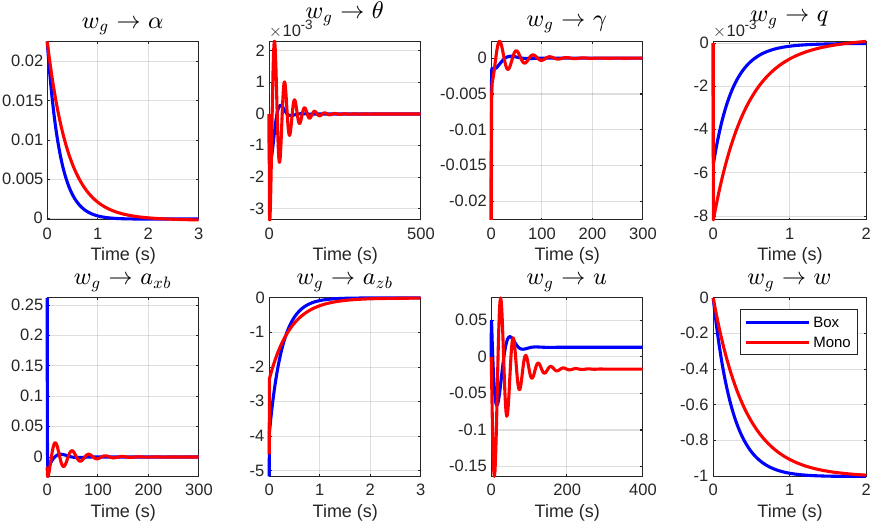}
\caption{B3V(blue), M3V(red).}
\flab{longi_step3}
\end{subfigure}
\caption{Time response to a step gust for longitudinal dynamics.}
\flab{longi_step}
\end{figure}

The eigenvalues for the lateral dynamics are shown in \fig{lat_evs}. We observe all the eigenvalues for all the configurations are real. Therefore, we do not anticipate any oscillations in the system's natural response. Also, we do not see significant differences between 6V and 62V configurations and expect their time responses to be similar. However, there is a noticeable difference between B3V and M3V configurations. We also observe there is a significant difference between 3V and 6V configurations. The time responses in \fig{lat_step} support these observations, which were obtained by applying step perturbation to $v$. In general, we observe no oscillations in the box-wing's time response, and it reaches a steady state faster than the mono-wing. 

\begin{figure}[h!]
\begin{subfigure}{\textwidth}
\centering

\includegraphics[width=0.7\textwidth]{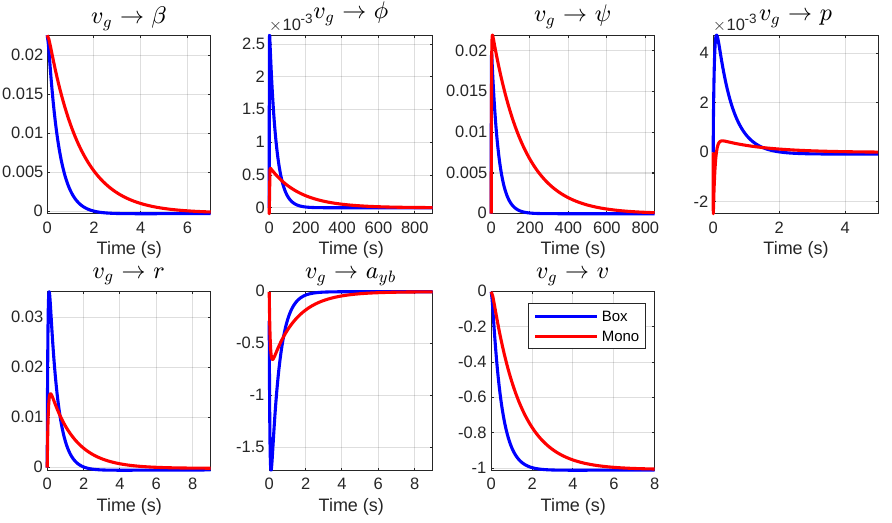}

\caption{B6V(blue), M6V(red)}
\flab{lat_step1}
\end{subfigure}
\begin{subfigure}{\textwidth}
\centering

\includegraphics[width=0.7\textwidth]{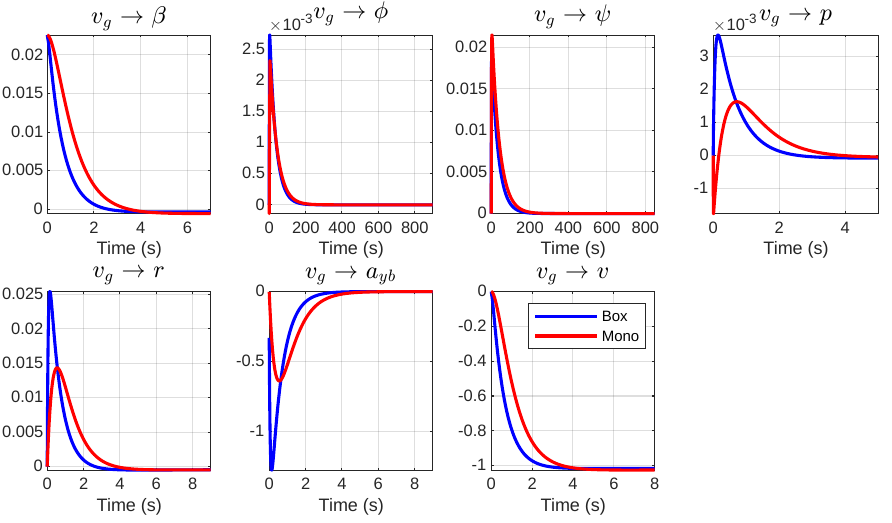}
\caption{B3V(blue), M3V(red)}
\flab{lat_step3}
\end{subfigure}
\caption{Time response to a step gust for lateral dynamics.}
\flab{lat_step}
\end{figure}

\section{Conclusions}
\label{sec:conclusion}
The aerodynamic analysis of a conventional mono-wing design was presented in comparison to a box-wing design at the small UAV scale. Comparing the mono-wing and box-wing designs proved non-trivial, as a non-conventional design like a closed-wing box-wing aircraft entails distinct parameter definitions. The aerodynamic characteristics of a box-wing were assessed against those of a mono-wing by altering geometric parameters of the chord, area, and aspect ratios, which yielded different Reynolds numbers of operation. 

The findings revealed that mono-wing configurations exhibited superior aerodynamic efficiency over box-wing designs across a broad spectrum of parameters, encompassing aspect ratio, velocity, and lift requirements. Nevertheless, box-wing configurations outperformed mono-wings at elevated velocities and with increased lift demands for a given aspect ratio. Generally, the box-wing design demonstrated greater favorability in situations where friction drag was less consequential than induced drag, highlighting its suitability for particular flight conditions.

With respect to flight dynamics, low aspect ratio box-wing configurations displayed heightened gust tolerance in both longitudinal and lateral dynamics, contributing to augmented stability and maneuverability. Conversely, when evaluating high aspect ratio configurations, no marked differences were discerned between box-wing and mono-wing designs in terms of gust tolerance. These critical insights emphasize the necessity of carefully selecting the optimal wing configuration, taking into account the specific performance requirements and operational conditions of the aircraft.

In summary, box-wing configurations possess inherent advantages that contribute to improved flight performance in challenging, gusty environments. The higher aerodynamic efficiency of these designs enables enhanced control authority, allowing for precise maneuverability and stability during flight. Additionally, the reduced structural stress associated with box-wings is crucial for maintaining structural airworthiness, ensuring the safety and longevity of the aircraft. These benefits are particularly advantageous in emerging micro air vehicle (MAV) applications in urban air mobility. Applications such as aerial surveillance and the transportation of high-value, low-weight payloads, like medical supplies or sensitive equipment, require reliable and efficient aerial platforms. With their unique aerodynamic and structural properties, box-wing configurations are well-suited to address these demands, showcasing their potential to revolutionize aerial operations in urban environments and beyond.
\section*{Appendix}

\section{Aerodynamic force and moment coefficients}
\label{App1}

\begin{figure}[htpb]
\centering
\begin{subfigure}{0.4\linewidth}
\centering
\includegraphics[width=0.99\textwidth]{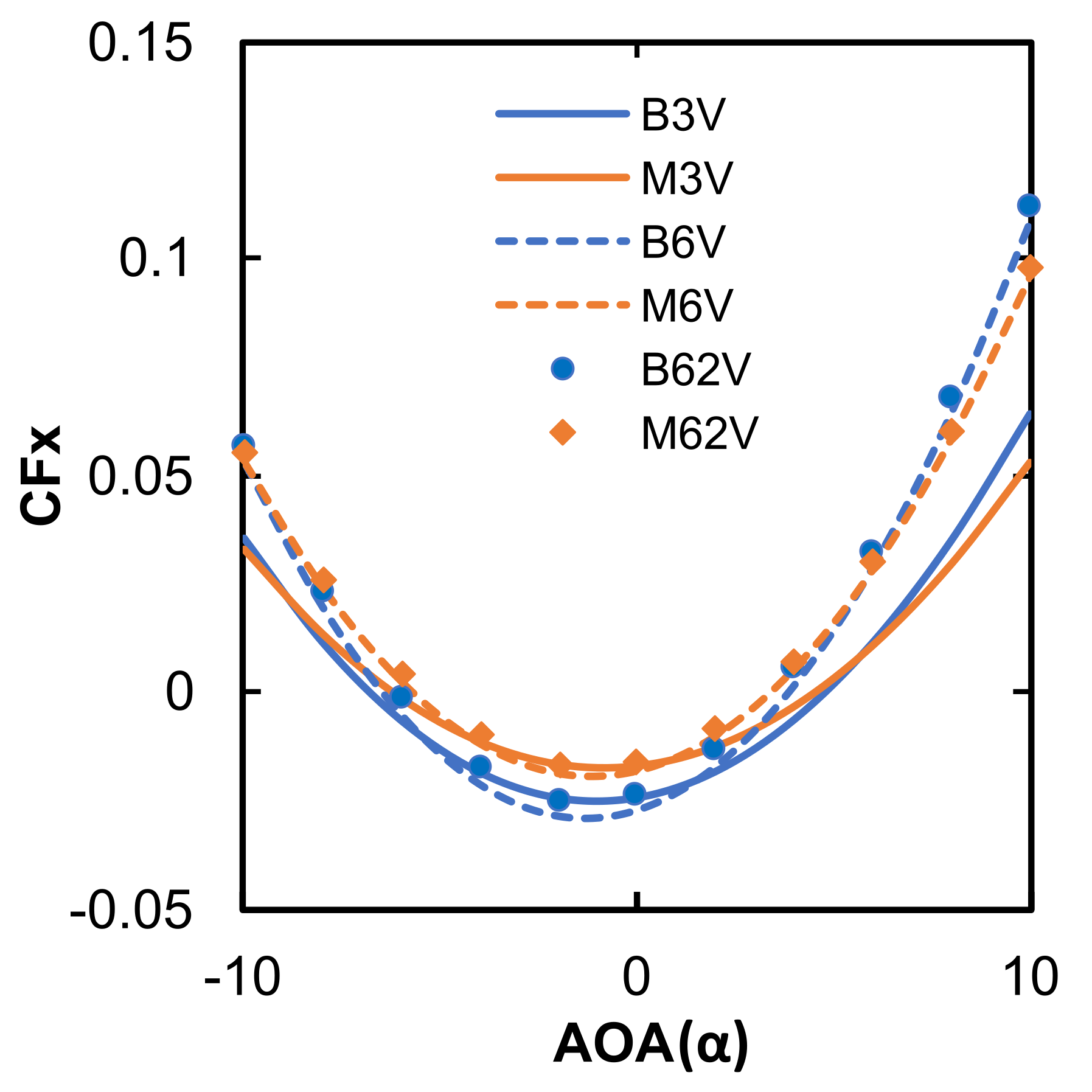}
\caption{}\label{fig:cfxvsa}
\end{subfigure}
\begin{subfigure}{0.4\linewidth}
\centering
\includegraphics[width=0.99\textwidth]{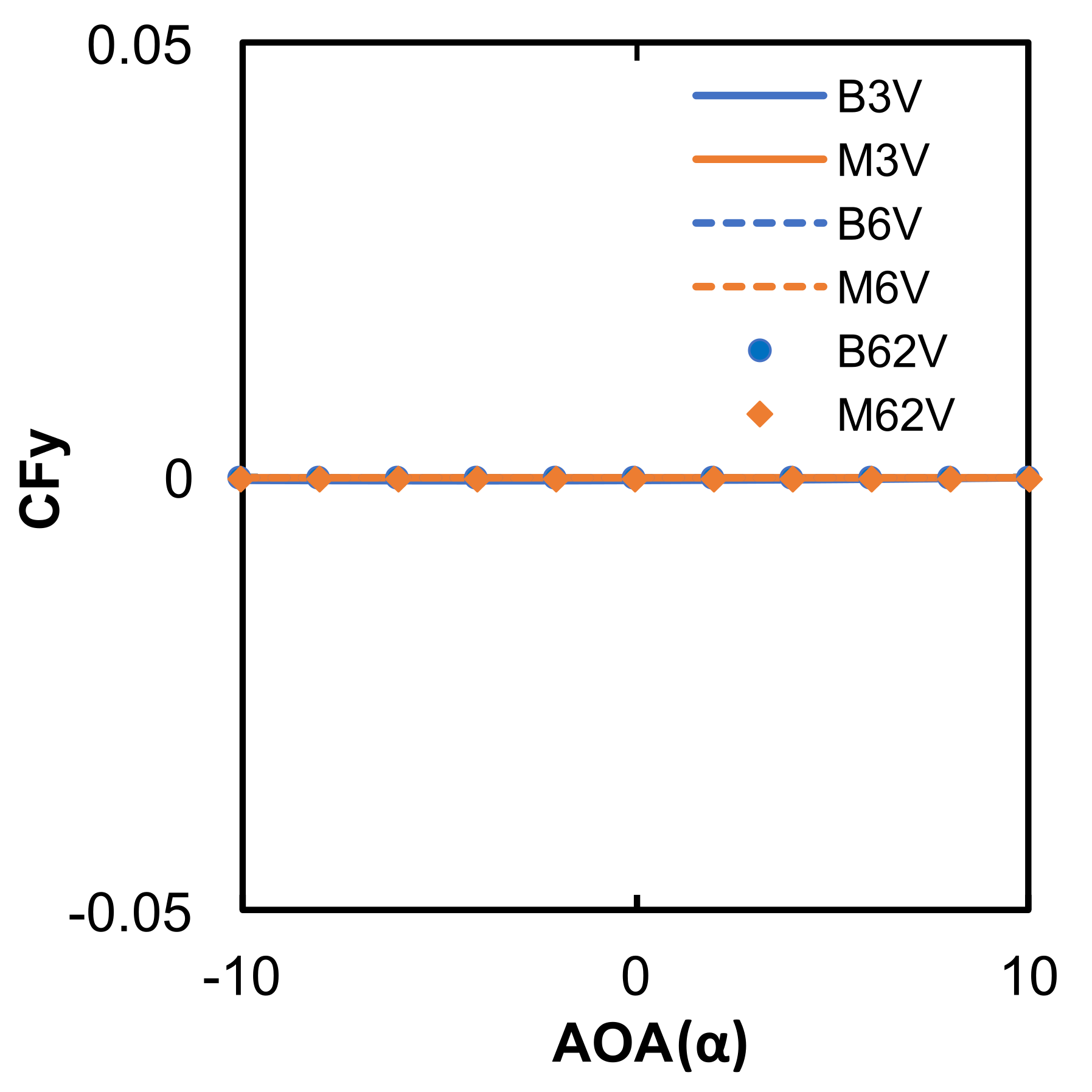}
\caption{}\label{fig:cfyvsa}
\end{subfigure}
\begin{subfigure}{0.4\linewidth}
\centering
\includegraphics[width=0.99\textwidth]{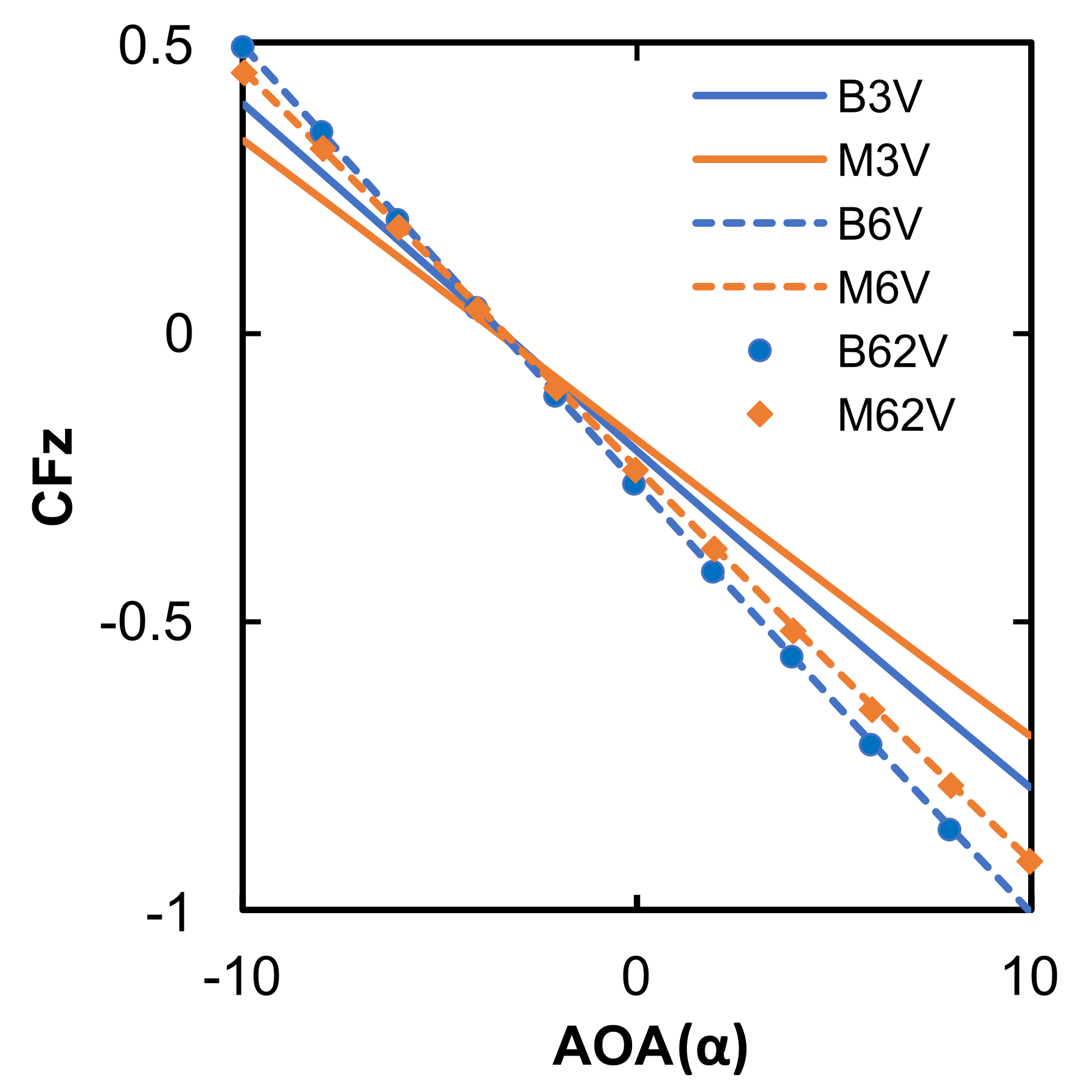}
\caption{}\label{fig:cfzvsa}
\end{subfigure}
\begin{subfigure}{0.4\linewidth}
\centering
\includegraphics[width=0.99\textwidth]{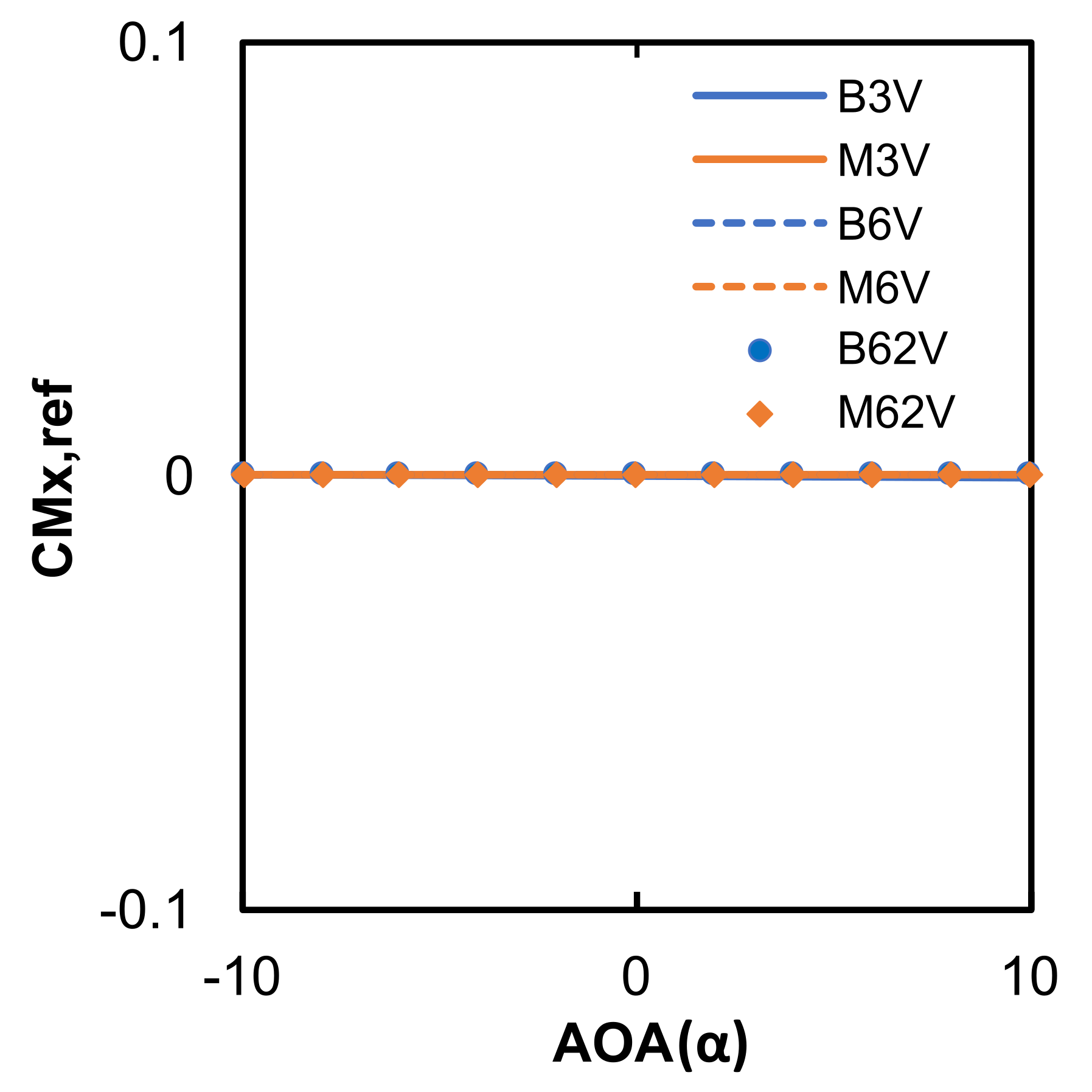}
\caption{}\label{fig:cmxvsa}
\end{subfigure}
\begin{subfigure}{0.4\linewidth}
\centering
\includegraphics[width=0.99\textwidth]{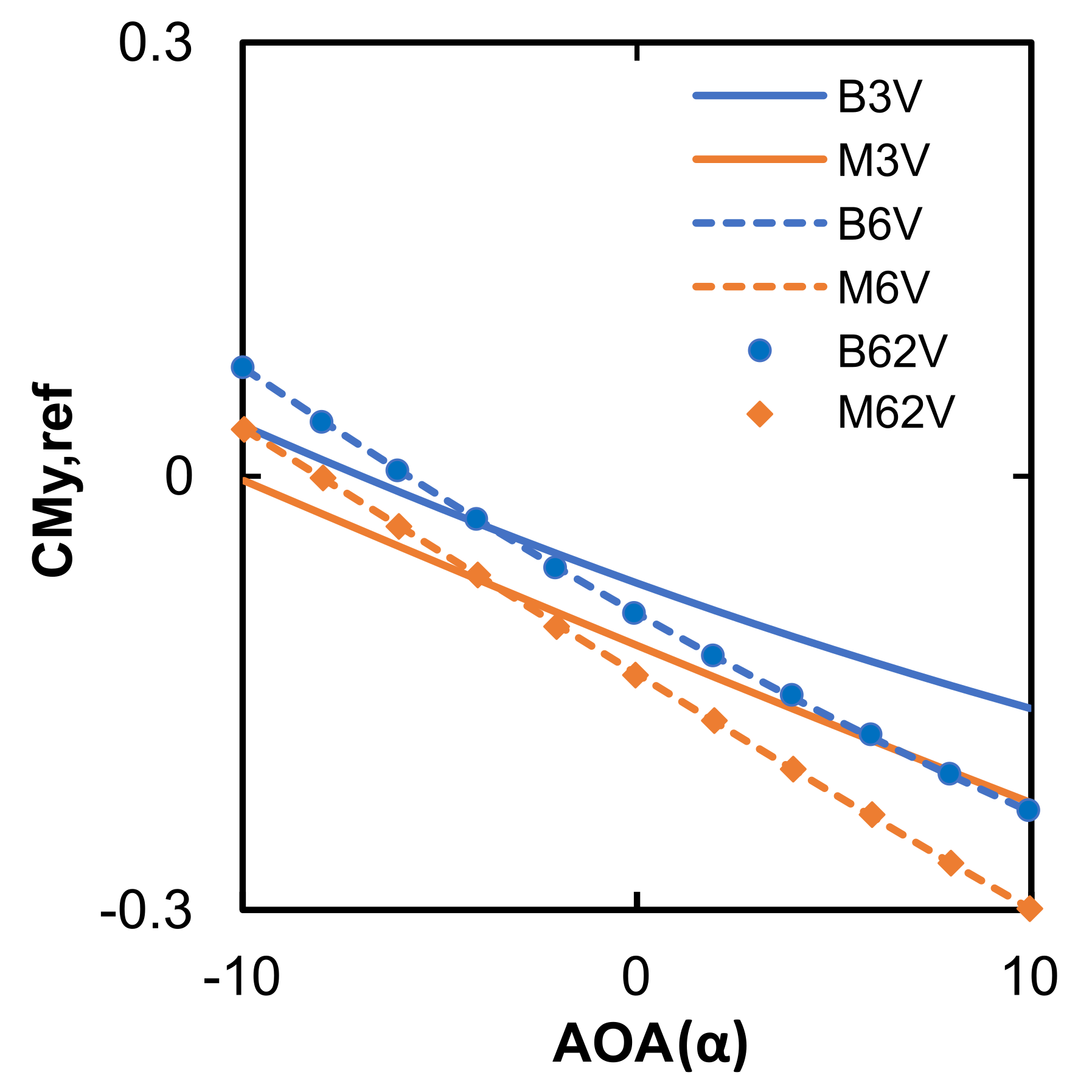}
\caption{}\label{fig:cmyvsa}
\end{subfigure}
\begin{subfigure}{0.4\linewidth}
\centering
\includegraphics[width=0.99\textwidth]{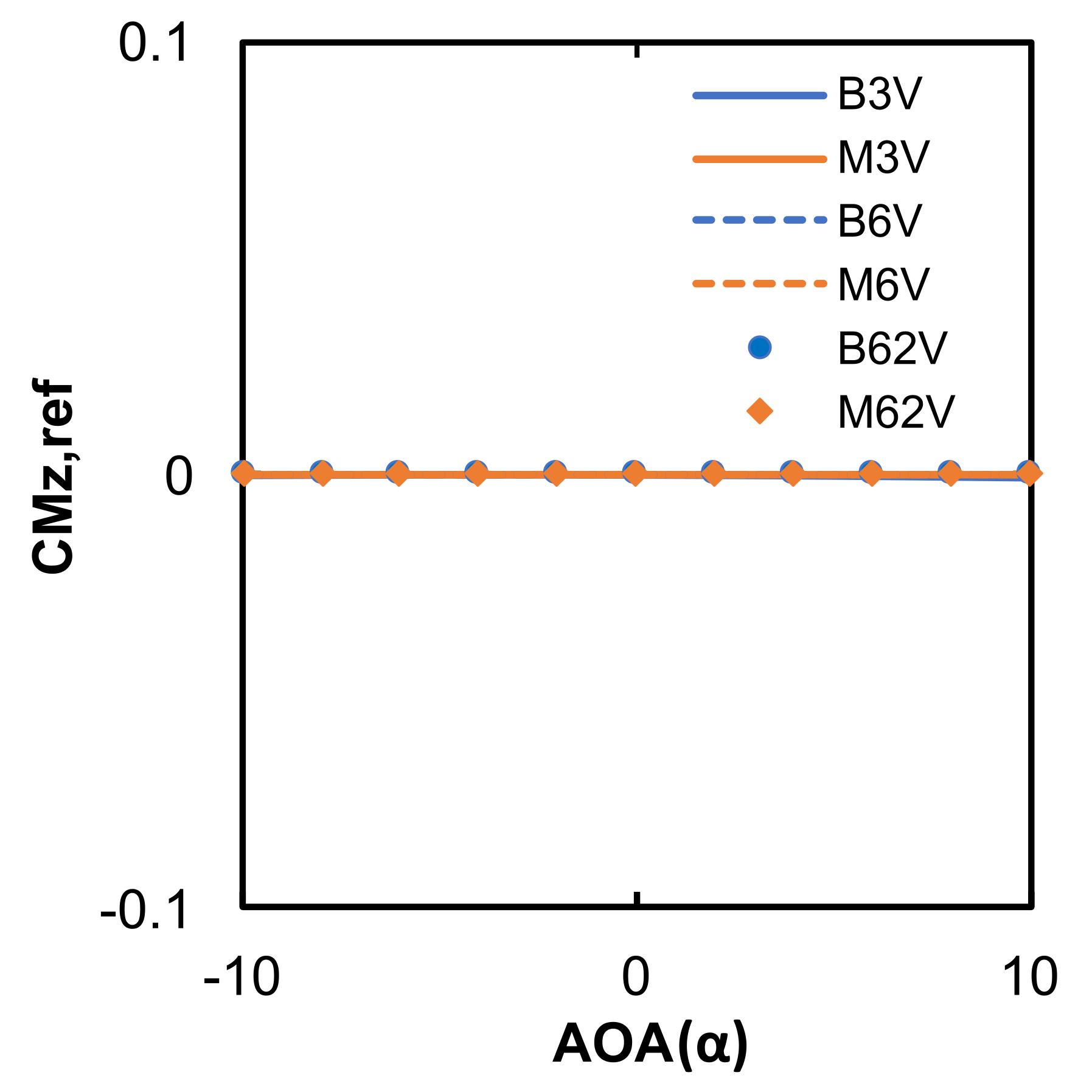}
\caption{}\label{fig:cmzvsa}
\end{subfigure}
\caption{$C_{Fx}$, $C_{Fy}$, $C_{Fz}$, $C_{Mx,\mathrm{ref}}$, $C_{My,\mathrm{ref}}$, $C_{My,\mathrm{ref}}$ at reference point vs. $\alpha$ at $\beta = 0$}
\label{fig:cf_cm_vs_a}       
\end{figure}
\begin{figure}[htpb]
\centering
\begin{subfigure}{0.4\linewidth}
\centering
\includegraphics[width=0.99\textwidth]{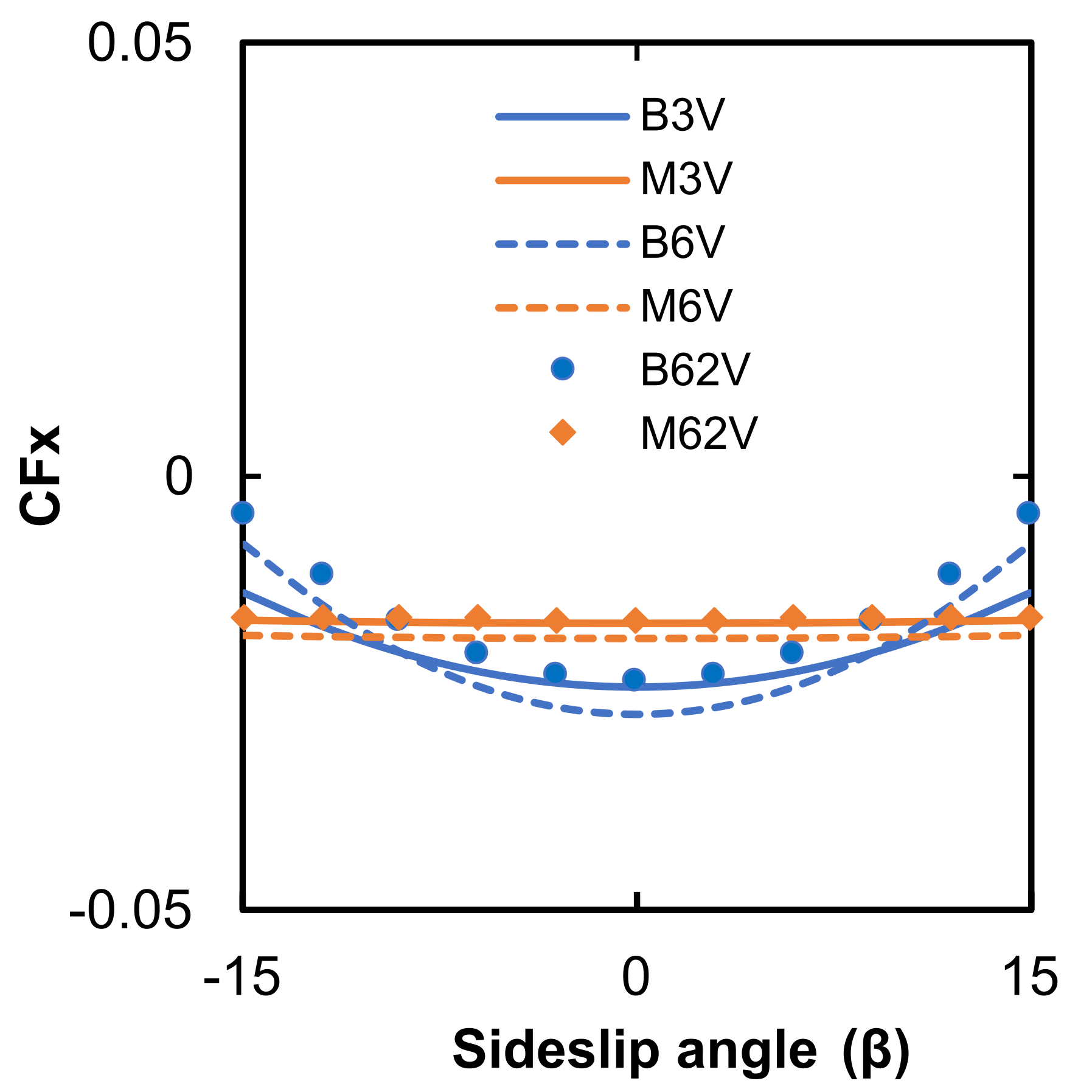}
\caption{}\label{fig:cfxvsb}
\end{subfigure}
\begin{subfigure}{0.4\linewidth}
\centering
\includegraphics[width=0.99\textwidth]{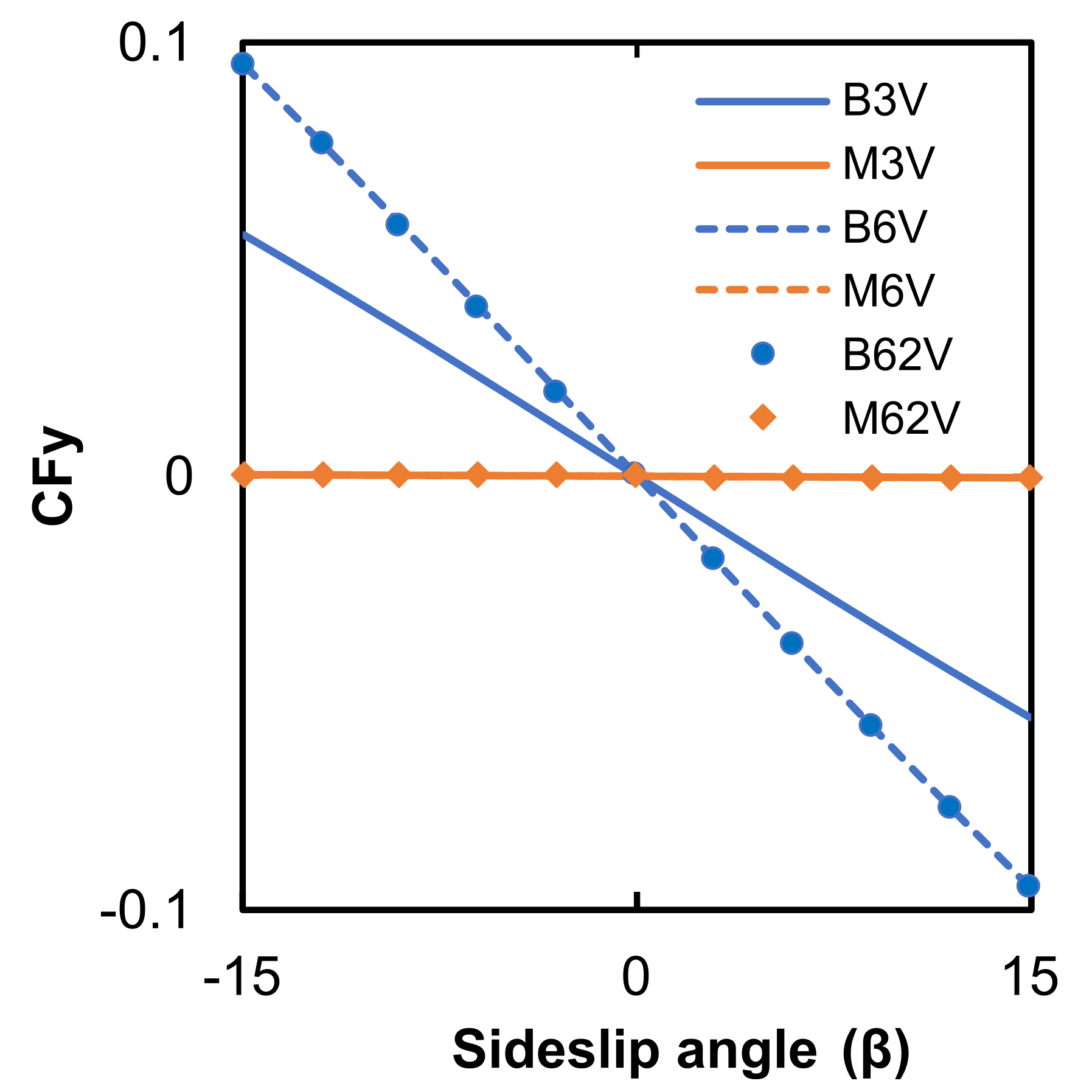}
\caption{}\label{fig:cfyvsb}
\end{subfigure}
\begin{subfigure}{0.4\linewidth}
\centering
\includegraphics[width=0.99\textwidth]{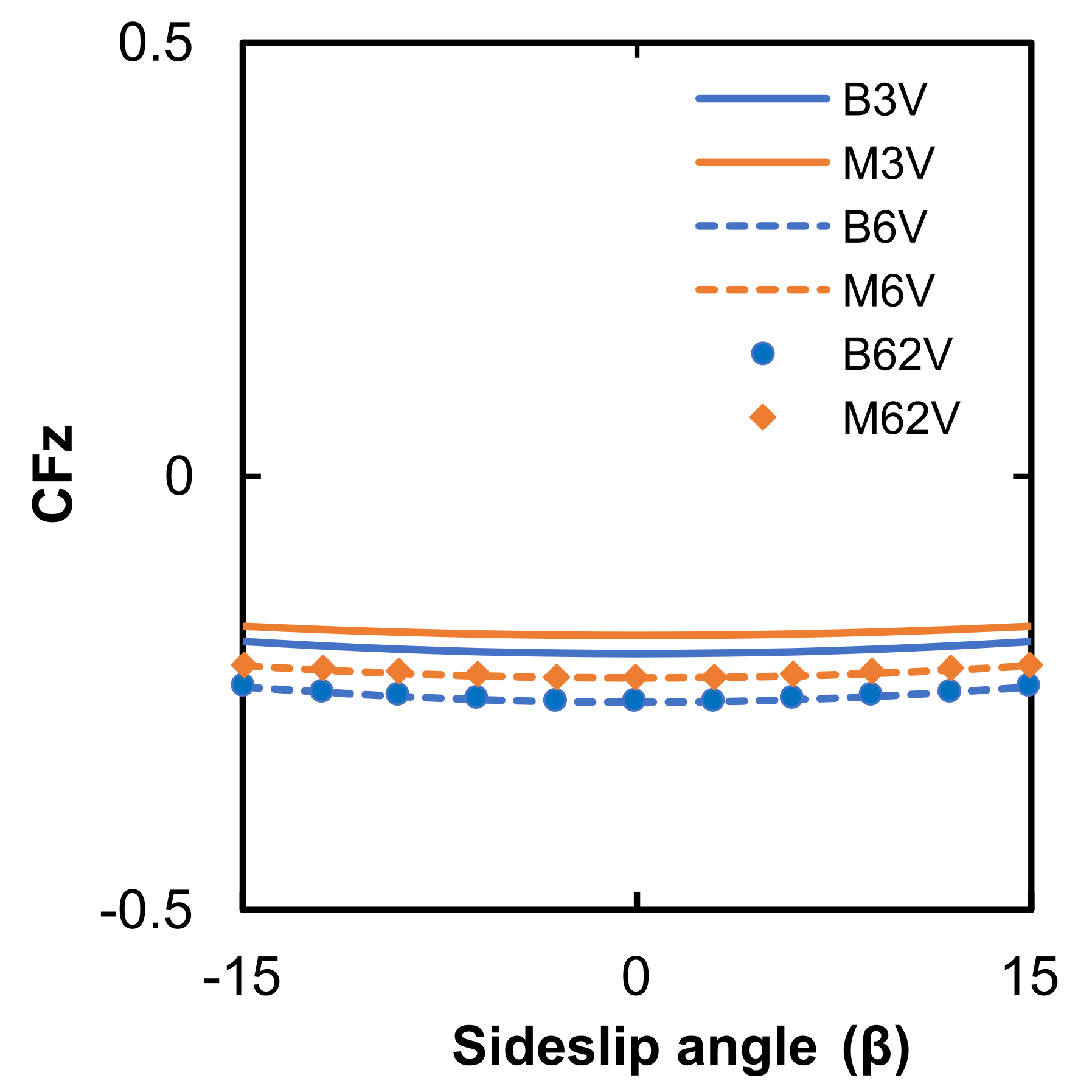}
\caption{}\label{fig:cfzvsb}
\end{subfigure}
\begin{subfigure}{0.4\linewidth}
\centering
\includegraphics[width=0.99\textwidth]{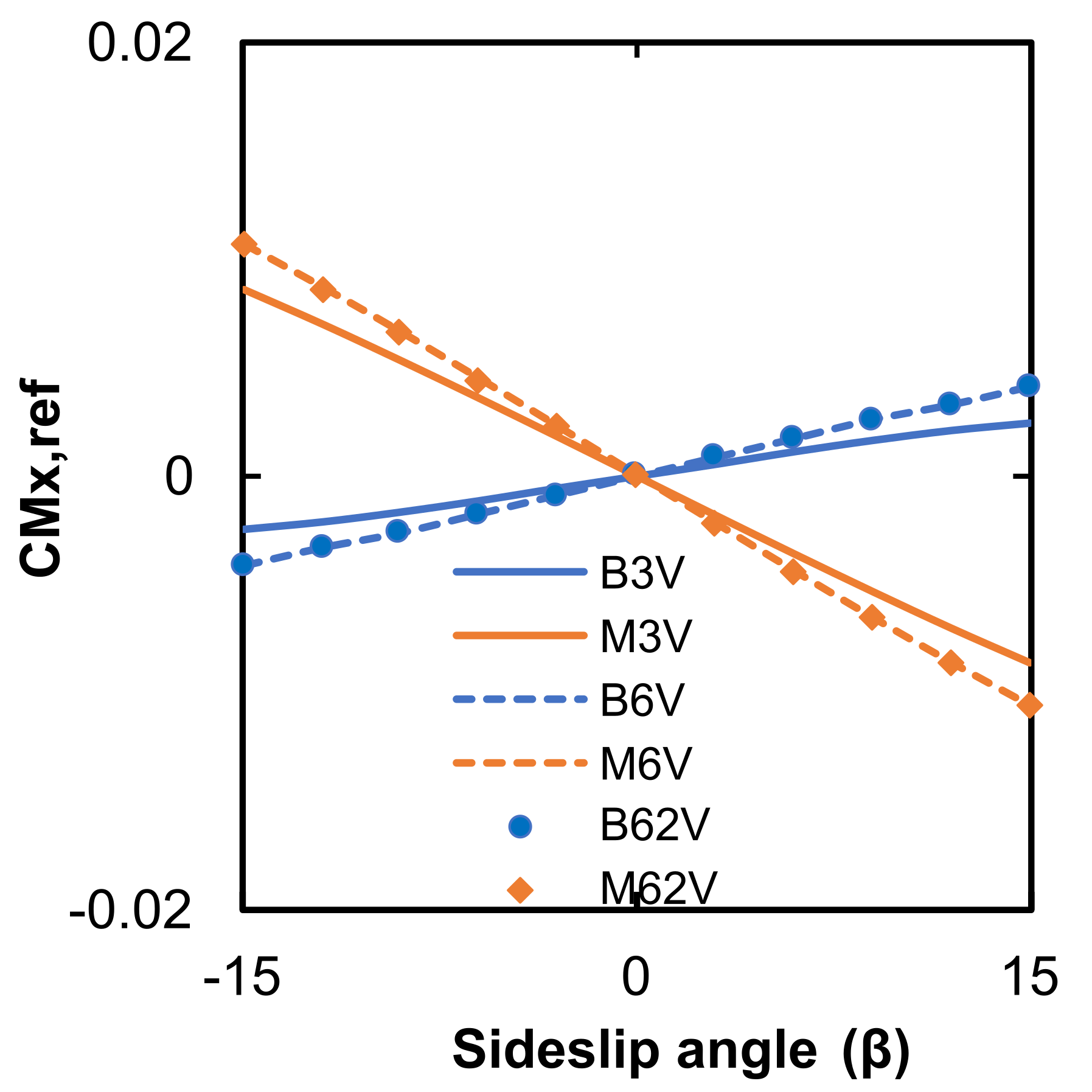}
\caption{}\label{fig:cmxvsb}
\end{subfigure}
\begin{subfigure}{0.4\linewidth}
\centering
\includegraphics[width=0.99\textwidth]{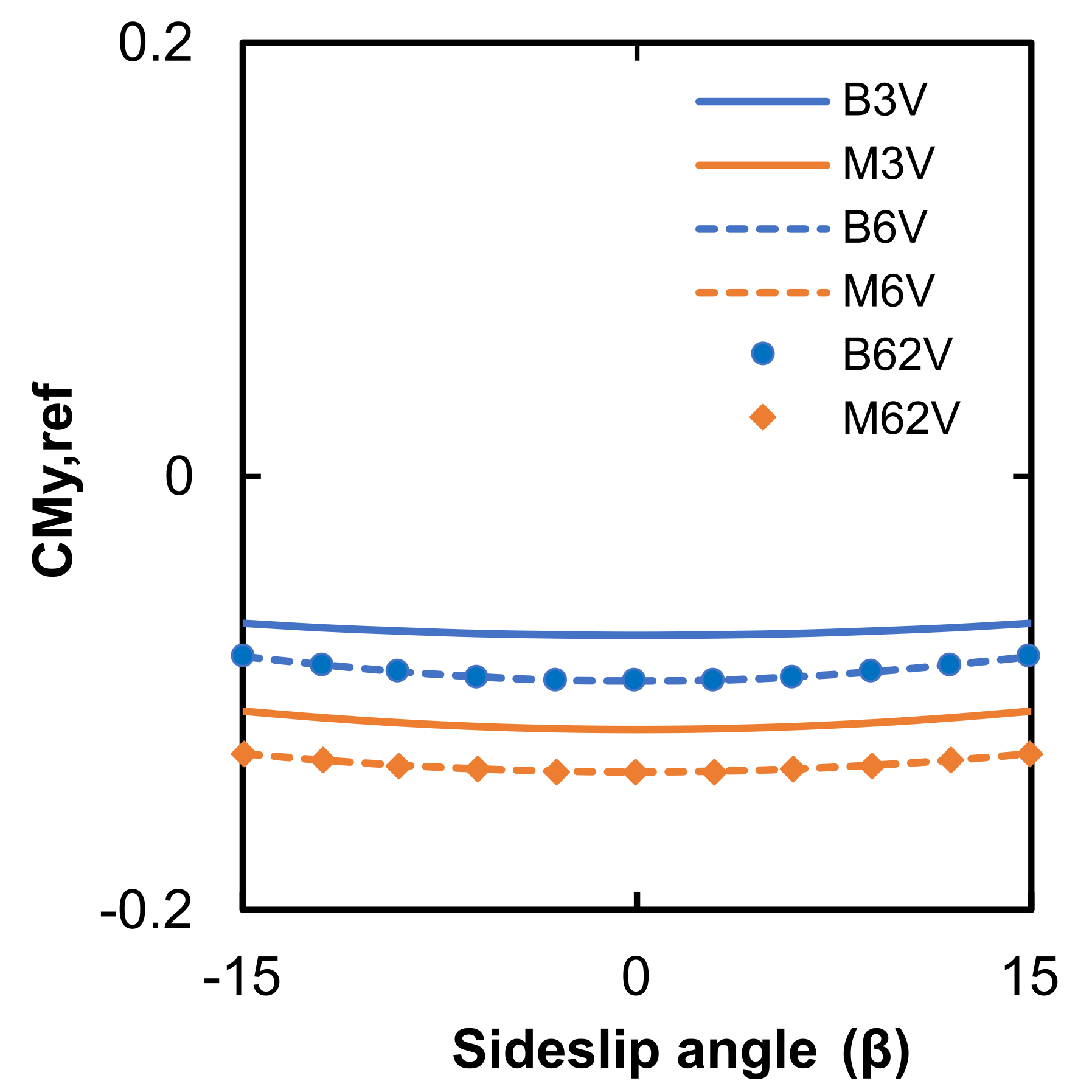}
\caption{}\label{fig:cmyvsb}
\end{subfigure}
\begin{subfigure}{0.4\linewidth}
\centering
\includegraphics[width=0.99\textwidth]{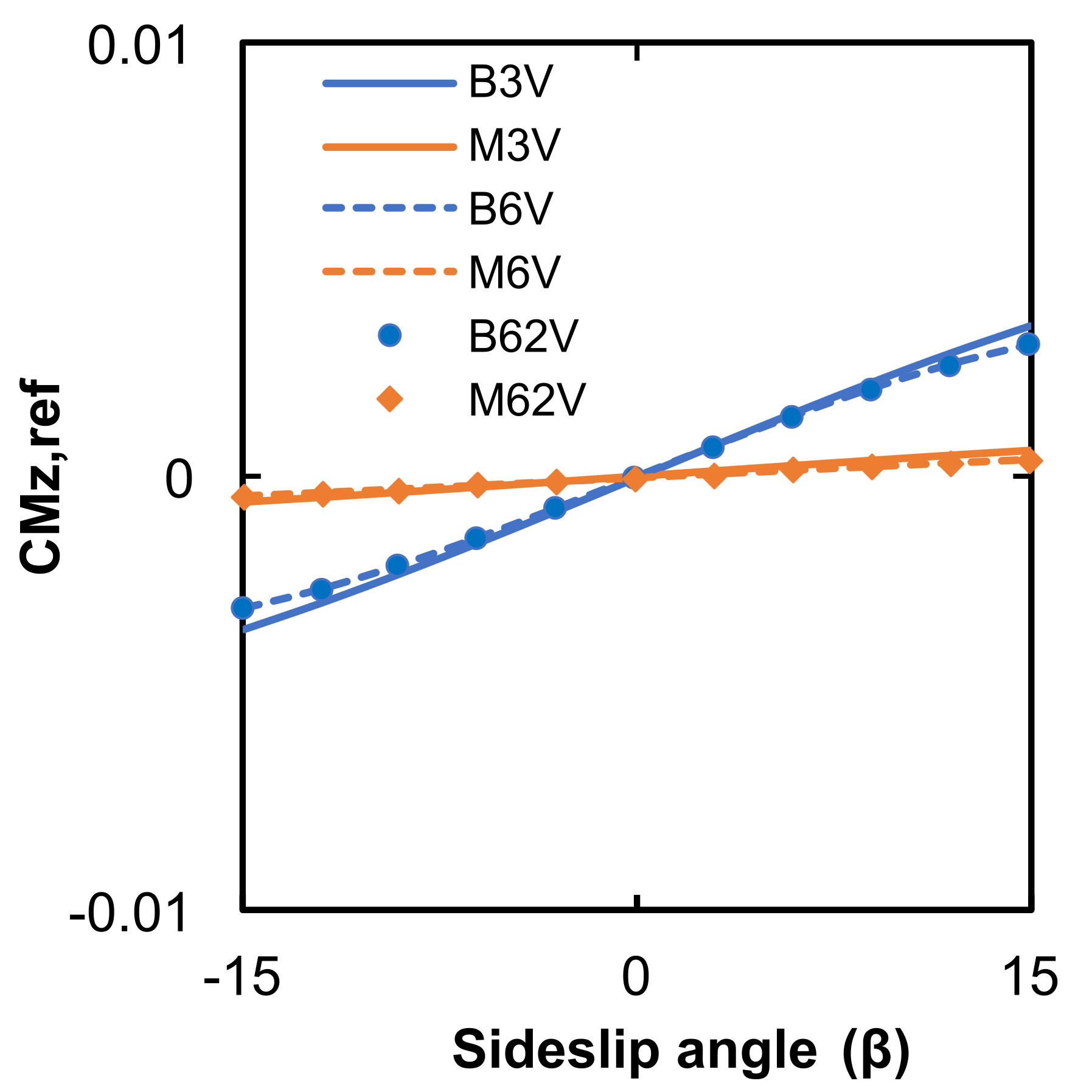}
\caption{}\label{fig:cmzvsb}
\end{subfigure}
\caption{$C_{Fx}$, $C_{Fy}$, $C_{Fz}$, $C_{Mx,\mathrm{ref}}$, $C_{My,\mathrm{ref}}$, $C_{My,\mathrm{ref}}$ at reference point vs. $\beta$ at $\alpha = 0$ }
\label{fig:cf_cm_vs_a_b}       
\end{figure}
\newpage
\section*{Acknowledgments}

The authors would like to acknowledge the computational facilities provided by the Department of Aerospace Engineering. We acknowledge the National Supercomputing Mission (NSM) for providing computing resources of `PARAM Shakti' at IIT Kharagpur, which is implemented by C-DAC and supported by the Ministry of Electronics and Information Technology (MeitY) and Department of Science and Technology (DST), Government of India.

\bibliography{sample}

\end{document}